\documentclass[useAMS,usenatbib]{mn2e}

\usepackage{graphicx}

\def\aV{\mbox{$\rm A_V$}}

\def\jh{\mbox{$(J-H)$}}

\def\jk{\mbox{$(J-K_s)$}}
\def\mMJ{\mbox{$(m-M)_J$}}
\def\mMo{\mbox{$(m-M)_O$}}
\def\ebv{\mbox{$E(B-V)$}}

\def\ejh{\mbox{$E(J-H)$}}

\def\rc{\mbox{$R_{\rm c}$}}

\def\rl{\mbox{$R_{\rm RDP}$}}
\def\rx{\mbox{$R_{\rm ext}$}}
\def\rt{\mbox{$R_{\rm t}$}}

\def\ms{\mbox{$M_\odot$}}
\def\ds{\mbox{$d_\odot$}}
\def\rs{\mbox{$R_\odot$}}
\def\dgc{\mbox{$R_{\rm GC}$}}
\def\xgc{\mbox{$X_{\rm GC}$}}
\def\ygc{\mbox{$Y_{\rm GC}$}}
\def\zgc{\mbox{$Z_{\rm GC}$}}

\def\jj{\mbox{$J$}}
\def\hh{\mbox{$H$}}
\def\ks{\mbox{$K_s$}}

\def\ns{\mbox{$N_{1\sigma}$}}
\def\no{\mbox{$N_{\rm obs}$}}
\def\nc{\mbox{$N_{\rm cl}$}}
\def\sFS{\mbox{$\rm\sigma_{FS}$}}
\def\fsU{\mbox{$FS_{\rm unif}$}}

\def\tdis{\mbox{$t_{\rm dis}$}}

\title[Investigation of 4 old infrared OCs]{Investigating the age and structure of the
infrared old open clusters LK\,1, LK\,10, FSR\,1521 and FSR\,1555}

\author[C. Bonatto and E. Bica]{C. Bonatto$^1$\thanks{E-mail: charles@if.ufrgs.br} and 
E. Bica$^1$\thanks{E-mail: bica@if.ufrgs.br}\\
$^1$ Departamento de Astronomia, Universidade Federal do Rio Grande do Sul\\ 
Av. Bento Gon\c{c}alves 9500, Porto Alegre 91501-970, RS, Brazil}

\begin{document}

\pagerange{\pageref{firstpage}--\pageref{lastpage}}

\maketitle

\label{firstpage}

\begin{abstract}
The combination of several mass-decreasing processes may critically affect the structure of 
open clusters (OCs), to the point that most dissolve into the field in a time-scale shorter 
than $\approx1$\,Gyr. Therefore, as is observed, old OCs should be sparse within the Galaxy. 
These mass-decreasing processes can only be quantifiably examined given the derivation of the 
fundamental parameters, age, distance and mass, for a sample of old OCs. In this paper we focus 
on 4 candidate old star clusters, namely, LK\,1, LK\,10, FSR\,1521, and FSR\,1555. The first 
two clusters, discovered by Le Duigou \& Kn\"odlseder, are projected towards the Cygnus 
Association, whilst the remaining two have been detected in the 4th quadrant as cluster 
candidates from stellar overdensities by Froebrich, Scholz \& Raftery. To analyse the target 
clusters we construct near infrared colour-magnitude diagrams (CMDs) and derive stellar radial 
density profiles (RDPs). The CMDs are constructed using 2MASS \jj, \hh, and \ks\ bands, and 
the intrinsic morphologies of the target OCs within these diagrams are revealed by applying a
field-star decontamination algorithm. Fundamental parameters are estimated with Padova 
isochrones built for the 2MASS filters. We derive extinctions to the objects within the range 
$3.4\le\aV\le8.9$, which makes them suitable for the near-infrared analysis, ages within 
$1.0 - 2.0$\,Gyr, and distances from the Sun within $1.4 - 4.5$\,kpc. These distances, in 
conjunction with the positions in the sky, place the present 4 OCs close to the solar circle 
($\la0.6$\,kpc). For LK\,10 our photometry reaches a depth $\approx3$\,mags below the main 
sequence turn off, from which we derive a relatively steep mass function slope ($\chi=2.4\pm0.4$) 
when compared to the Salpeter value ($\chi=1.35$). LK\,10 is a rather massive old OC, with a  
mass within $1360\le m(\ms)\le4400$, for stars in the observed magnitude range and the 
extrapolation to $0.08\,\ms$, respectively. The mass estimated in the restricted magnitude range 
for the remaining, more distant OCs is within $260\le m(\ms)\le380$. However, similarity with 
the CMD morphology and red clump of LK\,10 suggests that they may be as massive as LK\,10. The RDPs 
are well represented by a King-like function, except LK\,10, which has a central cusp suggesting 
post-core collapse. Structurally, LK\,1, FSR\,1521, and FSR\,155 are similar to a sample of nearby 
OCs of comparable age.  
\end{abstract}

\begin{keywords}
{\em (Galaxy:)} open clusters and associations: general; {\em (Galaxy:)} open clusters and
associations: individual:LK\,1, LK\,10, FSR\,1521, and FSR\,1555.
\end{keywords}

\section{Introduction}
\label{Intro}

The majority of the Galactic open clusters (OCs), the less massive ones in particular, do not  
survive the $\approx1$\,Gyr age barrier (\citealt{Friel95}; \citealt{OldOCs}, and references therein).
These systems are affected by several processes that continually change the gravitational potential, 
such as mass loss associated with stellar evolution, mass segregation and evaporation, tidal interactions 
with the Galactic disc and bulge, and collisions with giant molecular clouds. As clusters age, these 
mechanisms accelerate the internal dynamical evolution, which leads to important changes in the structure 
and flat, eroded mass functions. In most cases, OCs have their stellar content completely dissolved in the 
Galactic field, or leave only poorly-populated remnants (\citealt{PB07} and references therein).

Ample evidence gathered on theoretical (e.g. \citealt{Spitzer58}; \citealt{LG06}), N-body (e.g. 
\citealt{BM03}; \citealt{GoBa06}; \citealt{Khalisi07}), and observational (e.g. \citealt{vdB57}; 
\citealt{Oort58}; \citealt{vHoerner58}; \citealt{Piskunov07}) grounds indicate that, near the Solar circle, 
the disruption-time scale (\tdis) is shorter than $\sim1$\,Gyr. The disruption-time scale around the Solar 
circle appears to depend on mass as $\tdis\sim M^{0.62}$ (\citealt{LG06}). Thus, for clusters with mass 
within $10^2 - 10^3\ms$, a disruption time of $75\la\tdis(Myr)\la300$ should be expected. In
addition, disruption processes 
are stronger for the more centrally located and lower-mass OCs (see \citealt{OldOCs} for a review on 
disruption effects and time-scales). Gyr-class OCs are found preferentially near the Solar circle 
and in the outer Galaxy (e.g. \citealt{Friel95}; \citealt{DiskProp}), where the frequency of potentially 
disrupting dynamical interactions with giant molecular clouds and the disc is lower (e.g. \citealt{Salaris04};
\citealt{Upgren72}). As an extreme case, inner ($\dgc\la150$\,pc) Galactic tidal fields can dissolve a massive 
star cluster in a time-scale as short as $\sim50$\,Myr (\citealt{Portegies02}).

What should be expected from the above scenario is that only a small fraction of the OCs survive the
Gyr age-barrier, with the successful ones spending most of their existences preferentially at large 
Galactocentric distances.
Indeed, present-day statistics show that of the $\approx1000$ OCs with known age listed in the
WEBDA\footnote{\em obswww.univie.ac.at/webda} database, 180 are older than 1\,Gyr,
and only 18 ($\approx2\,\%$) are older than 4\,Gyr (see also \citealt{OBB05a}; \citealt{OBB05b}).
Additionally, most of the OCs older than 1\,Gyr so far identified are located outside the Solar 
circle (see, e.g., the spatial distribution of OCs of different ages in Fig.~1 of \citealt{OldOCs}).

\begin{figure*}
\begin{minipage}[b]{0.50\linewidth}
\includegraphics[width=\textwidth]{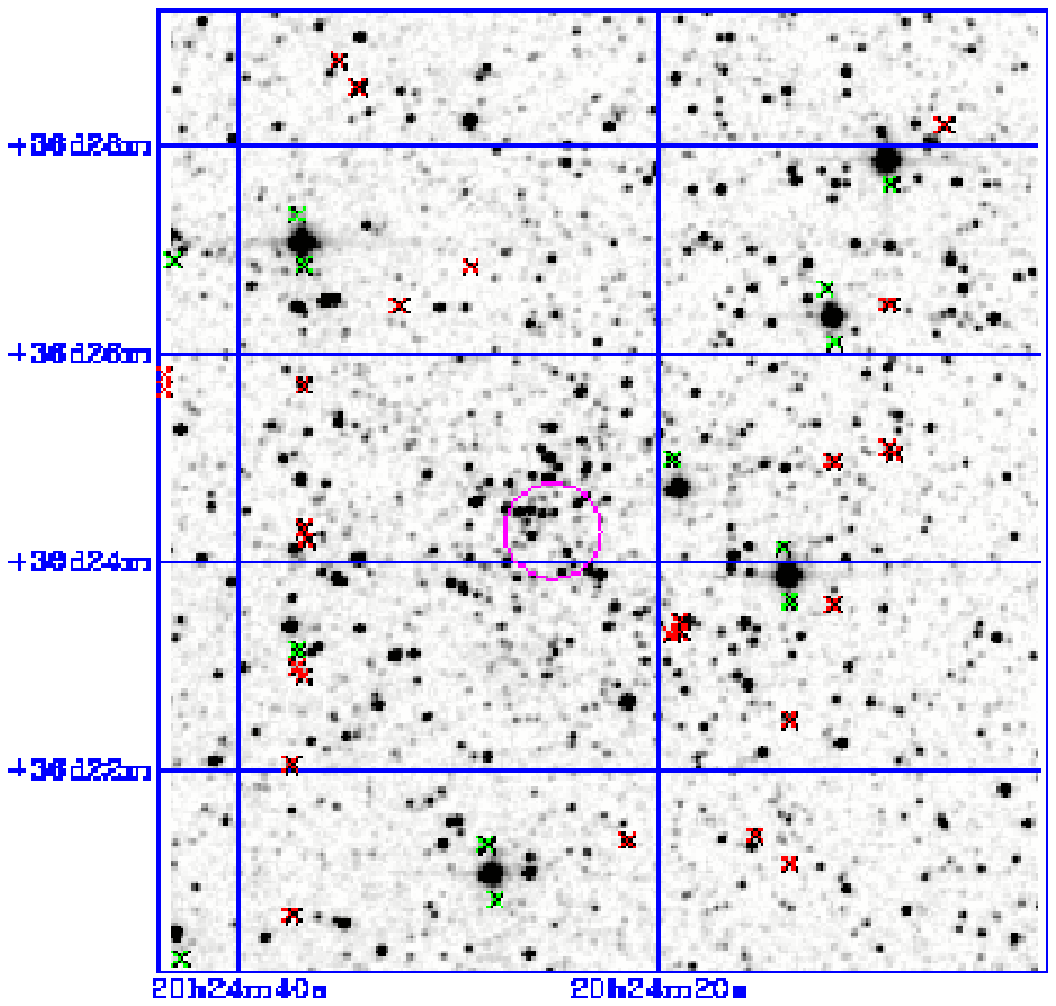}
\end{minipage}\hfill
\begin{minipage}[b]{0.50\linewidth}
\includegraphics[width=\textwidth]{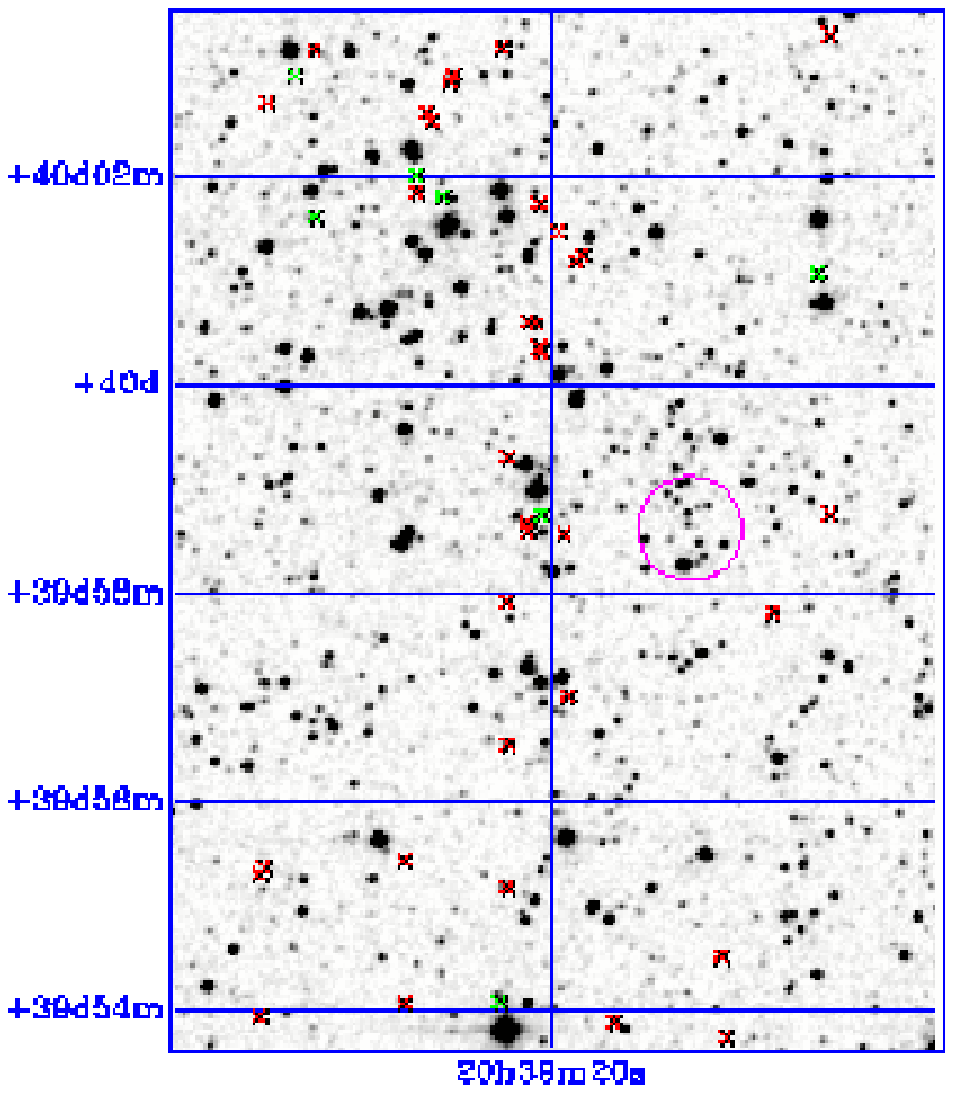}
\end{minipage}\hfill
\begin{minipage}[b]{0.50\linewidth}
\includegraphics[width=\textwidth]{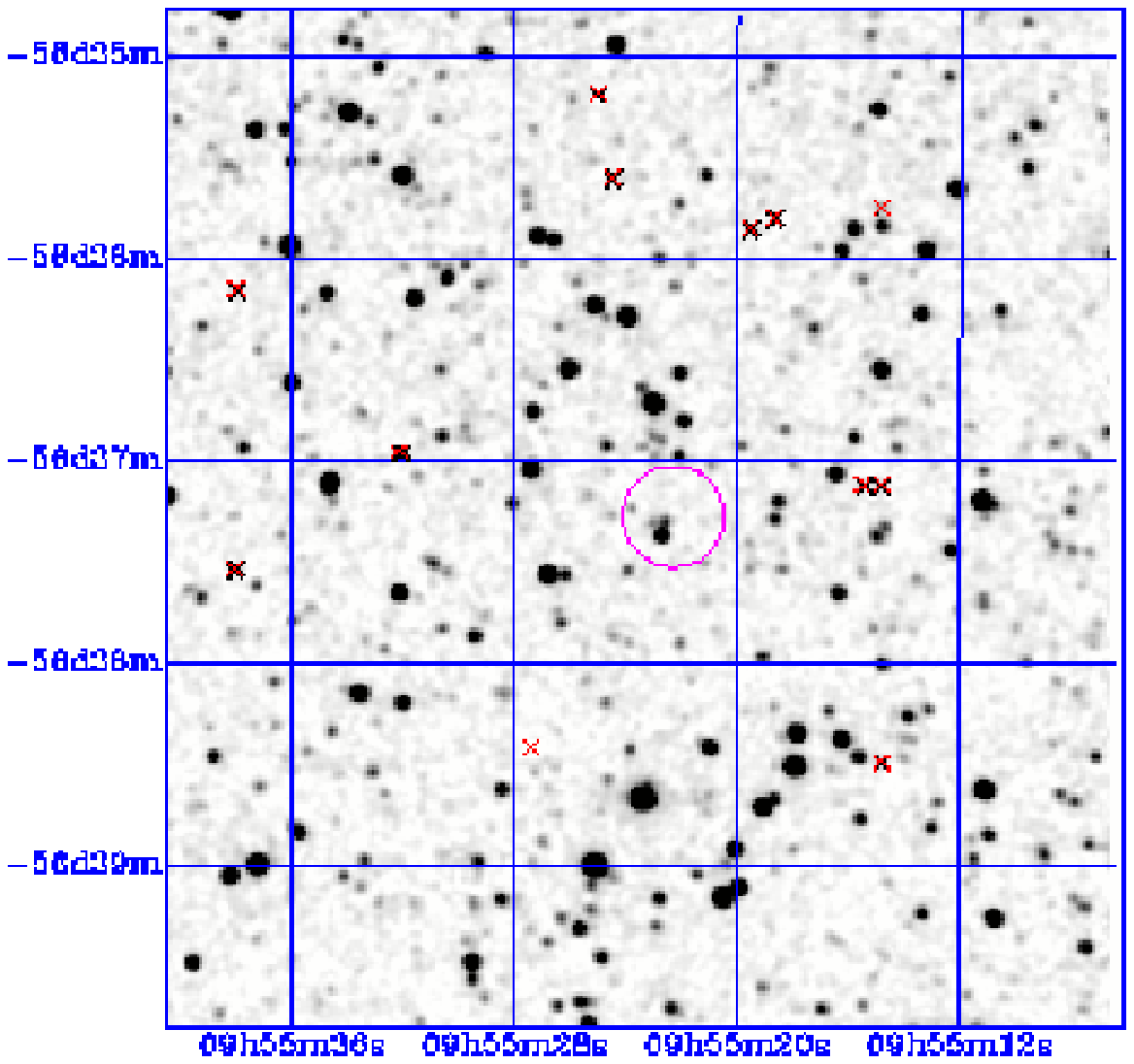}
\end{minipage}\hfill
\begin{minipage}[b]{0.50\linewidth}
\includegraphics[width=\textwidth]{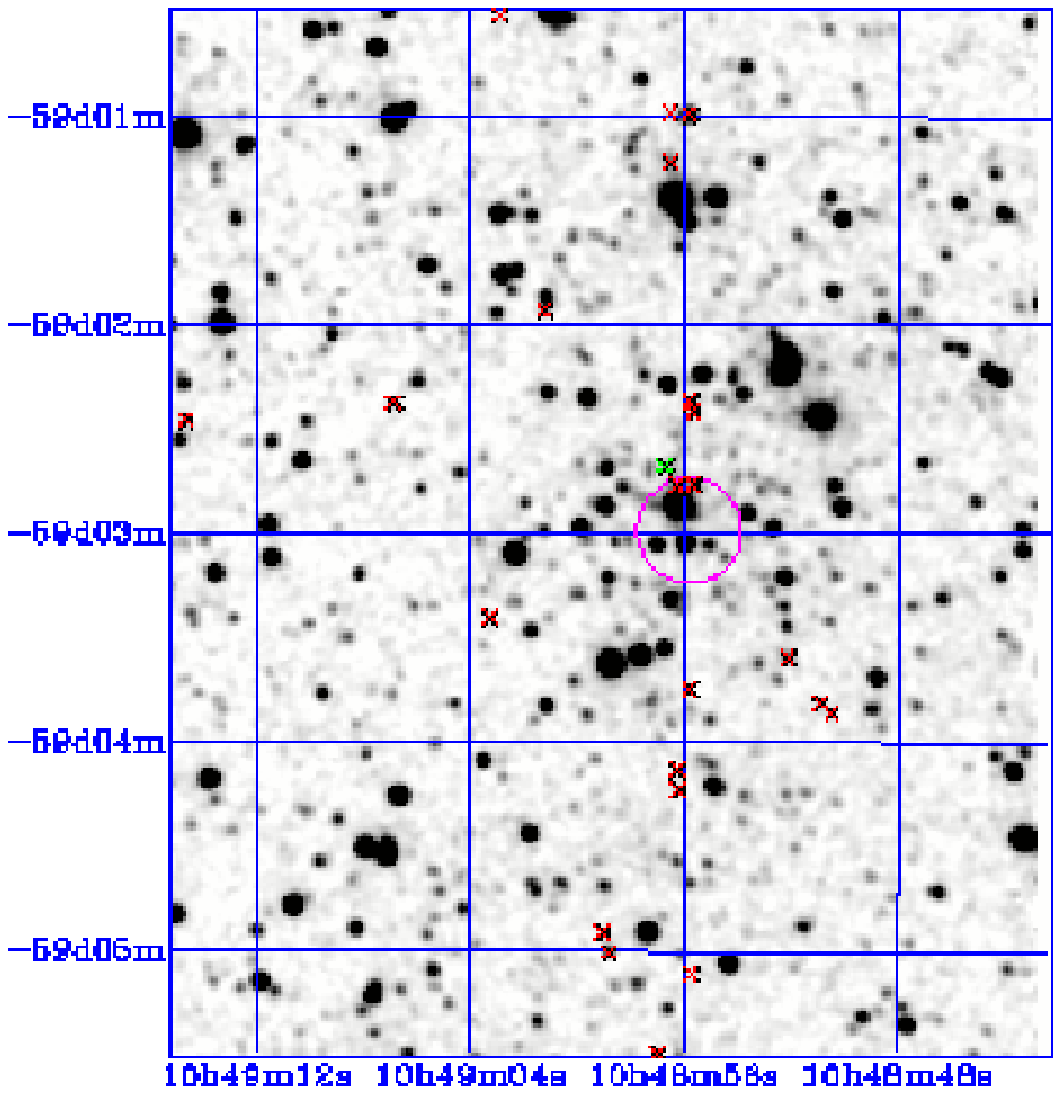}
\end{minipage}\hfill
\caption[]{Top panels: $5\arcmin\times5\arcmin$ 2MASS \ks\ images of LK\,1 (left) and LK\,10 (right). 
Bottom: same for FSR\,1521 (left) and FSR\,1555 (right). Images provided by the 2MASS Image Service. 
The small circle indicates the central coordinates (cols.~4 and 5 of Table~\ref{tab2}). Figure 
orientation: North to the top and East to the left. }
\label{fig1}
\end{figure*}

Given the aspects discussed above, a more thorough and wide-ranging exploration of the OC 
parameter space - the older ones in particular - is fundamental for a better understanding of
the dynamical process that affect the star clusters. In this sense, it is important to 
unambiguously establish the nature of as many old OCs as possible, followed by the derivation
of reliable astrophysical parameters. Such parameters, in turn, can be used in studies of star 
formation and evolution processes, dynamics and survival rates of star clusters, the geometry 
of the disc, among others. 

\begin{table}
\caption[]{Previous identifications of old clusters by our group}
\label{tab1}
\renewcommand{\tabcolsep}{0.4mm}
\renewcommand{\arraystretch}{1.25}
\begin{tabular}{lcl}
\hline\hline
Cluster& Age &~~~~~Reference\\
       &(Gyr)&         \\
\hline
FSR\,869&1.5& \citet{AntiCent}\\
FSR\,942&1.0& \citet{AntiCent}\\
FSR\,70 & $\ga5$& \citet{ProbFSR}\\
FSR\,124&1.0&\citet{ProbFSR}\\
ESO\,277SC1 (FSR\,1723)& 0.8&\citet{ProbFSR}\\
FSR\,1737 & $\ga5$& \citet{ProbFSR}\\
Ru\,101 (FSR\,1603) & 1.0 &\citet{F1603}\\
FSR\,1744& 1.0 &\citet{OldOCs}\\
FSR\,89  & 1.0 &\citet{OldOCs}\\
FSR\,31  & 1.1 &\citet{OldOCs}\\
Cz\,23 (FSR\,834)& 4.5&\citet{Cz23}\\
FSR\,1716&$\ga5$&\citet{Cz23}\\
FSR\,1415&  2.0&\citet{FSR1415}\\
Bica\,6 (BBS\,1)&1.0& \citet{PlaNeb}\\
\hline
\end{tabular}
\begin{list}{Table Notes.}
\item Bica\,6 (DAML02), located at $\alpha(2000)=05^h06^m20^s$ and 
$\delta(2000)=+39\degr09\arcmin50\arcsec$, is probably physically associated with the 
planetary nebula PK\,167-0.1.
\end{list}
\end{table}

\begin{table*}
\caption[]{Previous data and present results on the clusters}
\label{tab2}
\tiny
\renewcommand{\tabcolsep}{1.25mm}
\renewcommand{\arraystretch}{1.35}
\begin{tabular}{lccccccccccccccc}
\hline\hline
&\multicolumn{2}{c}{Literature}&&\multicolumn{9}{c}{This paper}\\
\cline{2-3}\cline{5-15}
Cluster&$\alpha(2000)$&$\delta(2000)$&&$\alpha(2000)$&$\delta(2000)$&$\ell$&$b$&Age&\aV&\ds&\dgc&\xgc&\ygc&\zgc\\
&(hms)&($^\circ\,\arcmin\,\arcsec$)
&&(hms)&($^\circ\,\arcmin\,\arcsec$)&($^\circ$)&($^\circ$)&(Gyr)&(mag)&(kpc)&(kpc)& (kpc)&(kpc)& (kpc)\\
(1)&(2)&(3)&&(4)&(5)&(6)&(7)&(8)&(9)&(10)&(11)&(12)&(13)&(14)\\
\hline
LK\,1&20:24:25&$+$36:24:18&&($\dagger$)&$+$36:24:30.0&75.24&$-0.69$&$1.0\pm0.2$&$8.9\pm0.3$
      &$4.0\pm0.4$&$7.3\pm0.2$&$-6.2\pm0.1$&$3.9\pm0.4$&$-0.05\pm0.01$\\
      
LK\,10&20:39:13&$+$39:58:37&&($\dagger$)&$+$39:58:43.2&79.84&$-0.92$&$1.0\pm0.1$&$8.5\pm0.3$
          &$1.4\pm0.1$&$7.1\pm0.1$  &$-7.0\pm0.1$&$1.4\pm0.1$&$-0.02\pm0.01$\\
          
FSR\,1521&09:55:23&$-$56:36:06&&09:55:22.3&$-$56:37:16.8&280.45&$-1.64$&$2.0\pm0.5$&$3.4\pm0.5$
      &$4.5\pm0.7$&$7.8\pm0.4$&$-6.4\pm0.1$&$-4.4\pm0.7$&$-0.13\pm0.02$\\
          
FSR\,1555&10:49:07&$-$59:03:17&&10:48:55.8&$-$59:02:58.9&287.75&$+0.16$&$1.5\pm0.5$&$3.7\pm0.3$
      &$4.1\pm0.2$&$7.1\pm0.1$&$-6.0\pm0.1$&$-3.9\pm0.2$&$0.01\pm0.01$\\
\hline
\end{tabular}
\begin{list}{Table Notes.}
\item Coordinates (cols.~2 and 3) of LK\,1 and LK\,10 from \citet{LK2002}; FSR\,1521 and FSR\,1555
from \citet{FSRcat}; Cols.~4-7: optimised coordinates; ($\dagger$): same value as in the literature;
Col.~9: reddening towards the cluster's central region (Sect.~\ref{age}). Col.~10: distance from the 
Sun. Col.~11: cluster Galactocentric distance for $\rs=7.2$\,kpc (\citealt{GCProp}). Cols.~12-14: 
coordinate components projected onto the Galactic plane. 
\end{list}
\end{table*}

In the present paper we focus on 4 candidate old OCs, namely, LK\,1, LK\,10, FSR\,1521, and
FSR\,1555. LK\,1 and LK\,10 are projected towards Cygnus and were found by \citet{LK2002}. An 
additional interesting point is to determine whether LK\,1 and/or LK\,10 are part of the Cygnus 
Association. FSR\,1521 and FSR\,1555 were found as candidate open clusters inferred from stellar 
overdensities by \citet{FSRcat}. 


This work employs near-IR \jj, \hh, and \ks\ 
photometry obtained from the 2MASS\footnote{The Two Micron All Sky Survey, All Sky data release
(\citealt{2mass1997}), available at {\em http://www.ipac.caltech.edu/2mass/releases/allsky/}} Point 
Source Catalogue (PSC). The spatial and photometric uniformity of 2MASS, which allow extraction 
of large surrounding fields that provide high star-count statistics, make it an excellent resource 
to gather photometric data on a broad variety of star clusters, the wide field ones in particular.
For this purpose we have developed quantitative tools to statistically disentangle cluster evolutionary 
sequences from field stars in colour-magnitude diagrams (CMDs), which are subsequently used to 
investigate the nature of star cluster candidates and to derive astrophysical parameters of the confirmed
clusters (e.g. \citealt{ProbFSR}). Basically, we apply {\em (i)} field-star decontamination to quantify 
the statistical significance of the CMD morphology, which is fundamental to deriving reddening, age, 
and distance from the Sun, and {\em (ii)} colour-magnitude filters, which are essential for intrinsic 
stellar radial density profiles (RDPs), as well as luminosity and mass functions (MFs). In particular,
the use of field-star decontamination in the construction of CMDs has proved to constrain the age and 
distance more than when working with the raw (observed) photometry, especially for low-latitude OCs 
(\citealt{DiskProp}).

This paper is organised as follows. In Sect.~\ref{RecAdd} we recall recent additions to the
known old OCs made by our group. Sect.~\ref{Target_OCs} contains basic properties and reviews
literature data (where available) on the present star cluster candidates. In Sect.~\ref{2mass} 
we present the 2MASS photometry, build CMDs, and apply the field-star decontamination algorithm.
In Sect.~\ref{age} we derive cluster fundamental parameters. Sect.~\ref{struc} describes cluster 
structure by means of stellar RDPs. In Sect.~\ref{MF} we provide estimates of cluster mass.  
In Sect.~\ref{Discus} we compare the structural parameters and dynamical state of the present 
clusters with those of a sample of nearby OCs, we also discuss effects of the location in
the Galaxy on their structure. Concluding remarks are given in Sect.~\ref{Conclu}.

\section{Recent additions to the known population of old OCs}
\label{RecAdd}

In recent years our group has been systematically analysing infrared clusters or candidates, 
establishing their nature and deriving cluster fundamental parameters with the 2MASS catalogue 
(e.g. \citealt{ProbFSR}). We make use of a field decontamination algorithm (described in 
Sect.~\ref{Decont_CMDs}) to statistically extract estimated cluster sequences from CMDs. Especially 
for crowded fields, the cluster sequence isolation requires some form of membership selection 
(e.g. \citealt{OldOCs}). Old OCs are intrinsically rarer than young clusters 
(Sect.~\ref{Intro}), but the 2MASS catalogue, coupled to the statistical tools that we have 
developed to deal with clusters and comparison fields allow the detection (and derivation 
of reliable astrophysical parameters) of older OCs (e.g. \citealt{OldOCs}; \citealt{F1603}).

Recently, \citet{FSRcat} provided a catalogue of star cluster candidates corresponding to 1021 stellar 
overdensities detected in the 2MASS database. This catalogue covers $|b|<20\degr$ and all Galactic 
longitudes, and has become an important source of new star clusters. Several followup studies have 
explored the FSR catalogue with different approaches, revealing new globular clusters, such as 
FSR\,1735 (\citealt{FSR1735}) and FSR\,1767 (\citealt{FSR1767}), and the probable GCs FSR\,584 
(\citealt{FSR584}) and FSR\,190 (\citealt{FSR190}). We show in Table~\ref{tab1} 14 recent 
identifications of old OCs made by our group. Three of them are optical objects, while the remaining
11 are infrared ones. Compared to the DAML02\footnote{http://www.astro.iag.usp.br/~wilton/ } optical 
catalogue of OCs, our results increased the known sample of old OCs by $\approx5\%$.

In the present paper we merge the old OCs in Table~\ref{tab1} with the classical
optical old OCs (WEBDA) for comparison purposes with the 4 objects dealt with in this
paper. This analysis is given in Sect.~\ref{Discus}.

\section{The target clusters}
\label{Target_OCs}

\citet{LK2002} provided a list of 17 star clusters and candidates in the Cygnus direction. As they point 
out, 12 of these objects (hereafter designated by LK) had already been found by \citet{DB2001}. Most of 
the LK objects appear to be embedded clusters, but some of them, like LK\,1 and LK\,10, have features 
typical of older clusters. With 2MASS photometry, they were able to estimate some cluster parameters.
For LK\,1 they derived a radius (containing 90\% of the stars) $R_{90}=3\farcm0$, the distance 
modulus $DM=11.0$, the absorption in the K band $A_K=1.0 - 2.0$, and a mass within $M=1100-3130\,\ms$; 
they suggested that LK\,1 may be a rather evolved cluster. As for LK\,10, they found $R_{90}=6\farcm7$, 
$DM=11.0$, $A_K=0.6 - 1.2$, $M=1010-4100\,\ms$, a rather steep mass function (MF: 
$\phi(m)\propto m^{-(1+\chi)}$) slope $\chi=1.98\pm0.25$, and suggested that it is quite evolved. 
Near-infrared 2MASS \ks\ images of LK\,1 and LK\,10, covering $5\arcmin\times5\arcmin$ fields, are shown 
in Fig.~\ref{fig1} (top panels).

FSR\,1521 was classified by \citet{FSRcat} as a highly probable star cluster candidate. They derived 
the core and tidal radii (measured in 2MASS \hh\ images) $R_c^H=1\farcm6$ and $R_t^H=25\farcm3$, 
respectively. FSR\,1521 can be seen in the $5\arcmin\times5\arcmin$ \ks\ image shown in Fig.~\ref{fig1} 
(bottom-left panel).

FSR\,1555 was also classified as a highly probable star cluster candidate by \citet{FSRcat}, who
derived $R_c^H=1\farcm7$ and $R_t^H=11\farcm6$ for this object. FSR\,1555 is shown in the 
$5\arcmin\times5\arcmin$ \ks\ image shown in Fig.~\ref{fig1} (bottom-right panel). In general, 
the clusters in the present sample are very contaminated by field stars, which requires specific 
tools to analyse them.

Table~\ref{tab2} provides fundamental data on the objects, where the literature coordinates are given 
in cols.~2 and 3. However, when we built the RDPs based on these coordinates (Sect.~\ref{struc}), we 
noticed that, in all cases, 
the coordinates where the maximum stellar number-density occurs are slightly shifted with respect to 
the literature positions. Thus, hereafter we will refer as cluster coordinates those that maximise
the central stellar density (given in cols.~4-7). The age, central reddening, distance from the Sun, 
Galactocentric distance, and the components projected onto the Galactic plane derived in the present 
study (Sect.~\ref{age}) are given in Cols.~8 to 14.

\section{Colour-magnitude diagrams with 2MASS photometry}
\label{2mass}

Photometry in the 2MASS \jj, \hh, and \ks\ bands was extracted in circular fields of radius
\rx\ centred on the coordinates of the objects (Table~\ref{tab2}) by means of VizieR\footnote{\em
http://vizier.u-strasbg.fr/viz-bin/VizieR?-source=II/246}. \rx\ should be large enough to allow 
the determination of the background level (Sect.~\ref{struc}). In the present cases, $\rx=30\arcmin$ 
(FSR\,1521 and FSR\,1555), $\rx=40\arcmin$ (LK\,1), and $\rx=60\arcmin$ (LK\,10), which are considerably 
larger than the cluster radius (Sect.~\ref{struc} and col.~5 of Table~\ref{tab4}). Previous works with 
OCs in different environments (Sect.~\ref{Intro}) have shown that in the absence of a populous neighbouring 
cluster and significant differential absorption (Sect.~\ref{AlgoDescr}), wide extraction areas provide 
the necessary statistics for a consistent colour and magnitude characterisation of the field stars. For 
decontamination purposes, comparison fields were extracted within wide rings located beyond the cluster 
radii. As photometric quality constraint, the 2MASS extractions were restricted to stars {\em (i)} brighter 
than those of the 99.9\% Point Source Catalogue completeness limit\footnote{According to the 2MASS 
Level\,1 Requirement, at {\em http://www.ipac.caltech.edu/2mass/releases/allsky/doc/ }} in the cluster 
direction, and {\em (ii)} with errors in \jj, \hh, and \ks\ smaller than 0.3\,mag. The 99.9\% completeness 
limits refer to field stars, and depend on Galactic coordinates. For the present clusters, the fraction
of stars with \jj, \hh, and \ks\ uncertainties smaller than 0.06\,mag is $\approx80\%$. A typical distribution 
of uncertainties as a function of magnitude, for clusters projected towards the central parts of the Galaxy, 
can be found in \citet{BB07}. Reddening transformations use the relations $A_J/A_V=0.276$, $A_H/A_V=0.176$,
$A_{K_S}/A_V=0.118$, and $A_J=2.76\times\ejh$ (\citealt{DSB2002}), for a constant total-to-selective 
absorption ratio $R_V=3.1$. These ratios were derived from the extinction curve of \citet{Cardelli89}.

\begin{figure}
\resizebox{\hsize}{!}{\includegraphics{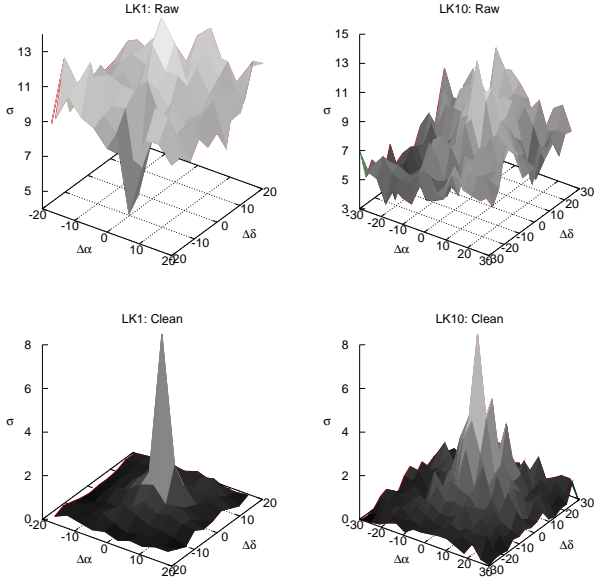}}
\caption[]{Stellar surface-density $\sigma(\rm stars\ arcmin^{-2})$ of LK\,1 (left panels)
and LK\,10 (right). The curves were computed for a mesh size of $3\arcmin\times 3\arcmin$, centred
on the coordinates in Table~\ref{tab2}. The observed (raw) and field-star decontaminated photometry
are shown in the top and bottom panels, respectively.}
\label{fig2}
\end{figure}

\begin{figure}
\resizebox{\hsize}{!}{\includegraphics{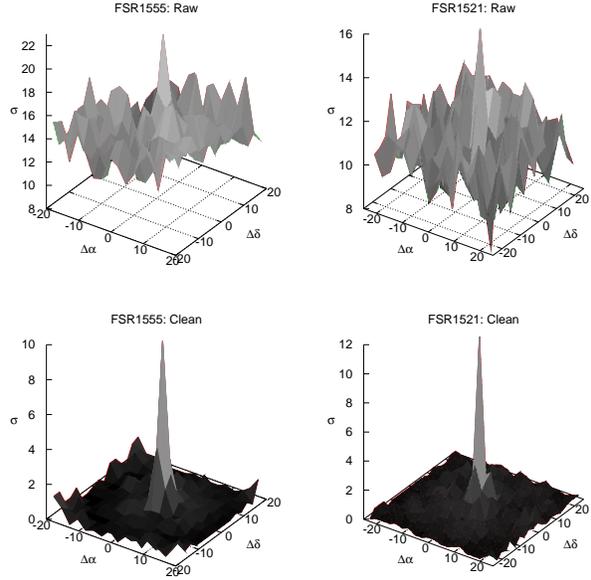}}
\caption[]{Same as Fig.~\ref{fig2} for FSR\,1555 (left) and FSR\,1521 (right).}
\label{fig3}
\end{figure}

\begin{figure}
\resizebox{\hsize}{!}{\includegraphics{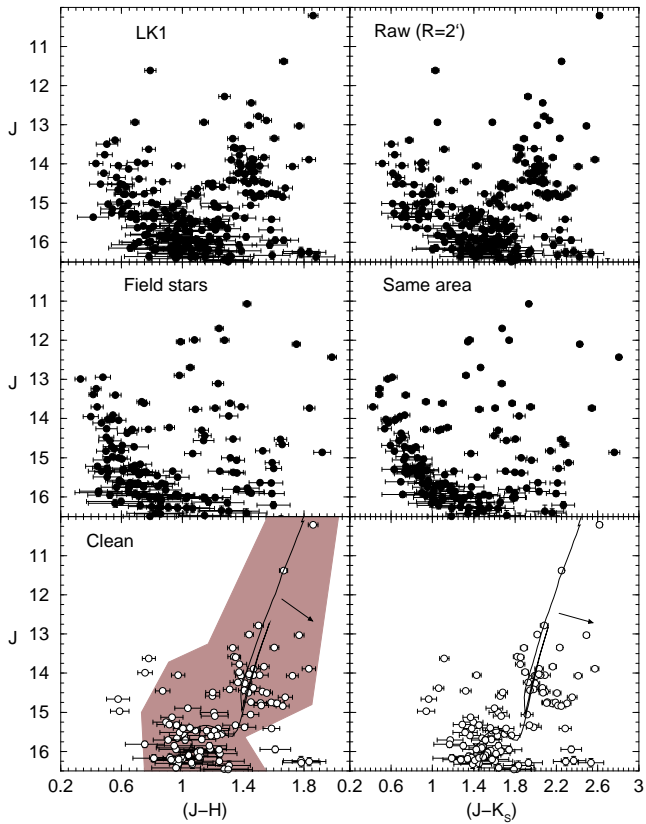}}
\caption[]{2MASS CMDs extracted from the $R<2\arcmin$ region of LK\,1. Top panels:
observed photometry with the colours $\jj\times\jh$ (left) and $\jj\times\jk$ (right). Middle:
equal-area ($29.93\arcmin<R<30\arcmin$) extraction from the comparison field, where the disc 
contamination is present. Bottom panels: decontaminated CMDs that suggest a relatively reddened
and distant MSTO, red clump, and giant branch typical of old OCs, fitted with the 1\,Gyr 
Solar-metallicity Padova isochrone. The shaded polygon corresponds to the colour-magnitude 
filter (Sect.~\ref{struc}). Arrows in the bottom panels show the reddening vector computed for 
$\aV=2$.}
\label{fig4}
\end{figure}

\begin{figure}
\resizebox{\hsize}{!}{\includegraphics{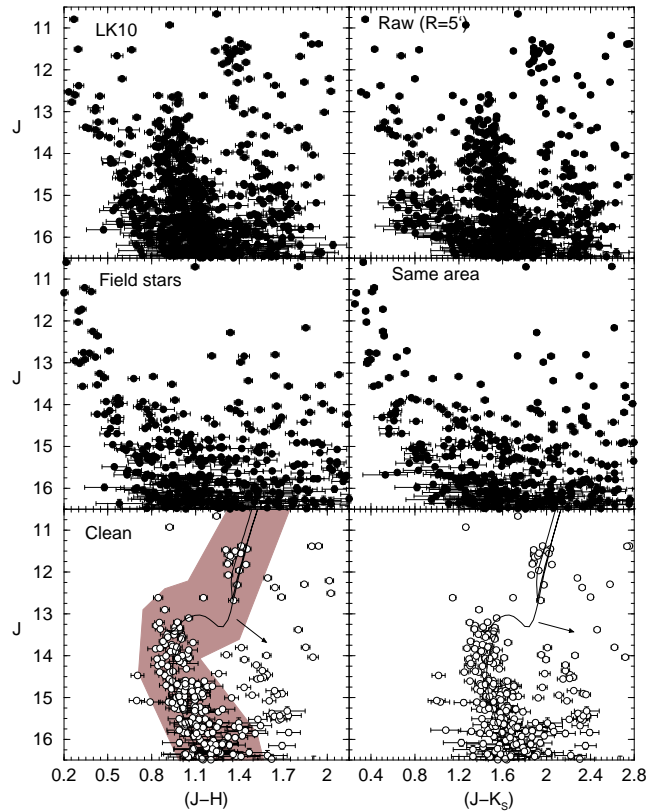}}
\caption[]{Same as Fig.~\ref{fig4} for the region $R<5\arcmin$ of LK\,10. The equal-area comparison 
field extraction was taken from the region $29.58\arcmin<R<30\arcmin$. A relatively populous giant 
clump and about 3.5 mag of the MS show up, especially in the decontaminated CMDs, denoting advanced 
age. The 1\,Gyr Solar-metallicity Padova isochrone is applied to the CMDs. Reddening vectors as 
in Fig.~\ref{fig4}.}
\label{fig5}
\end{figure}

\begin{figure}
\resizebox{\hsize}{!}{\includegraphics{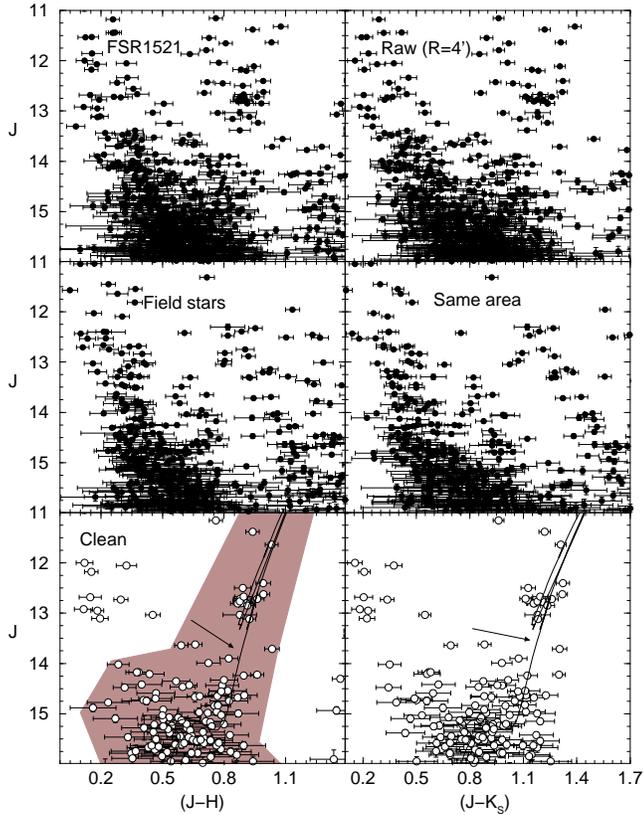}}
\caption[]{Same as Fig.~\ref{fig4} for the region $R<4\arcmin$ of FSR\,1521, with the equal-area 
comparison field extraction taken from $29.73\arcmin<R<30\arcmin$. The decontaminated CMDs suggest 
a giant clump and a red giant branch of an old distant OC. The age solution corresponds to the 
2\,Gyr isochrone. Reddening vectors as in Fig.~\ref{fig4}.}
\label{fig6}
\end{figure}

\begin{figure}
\resizebox{\hsize}{!}{\includegraphics{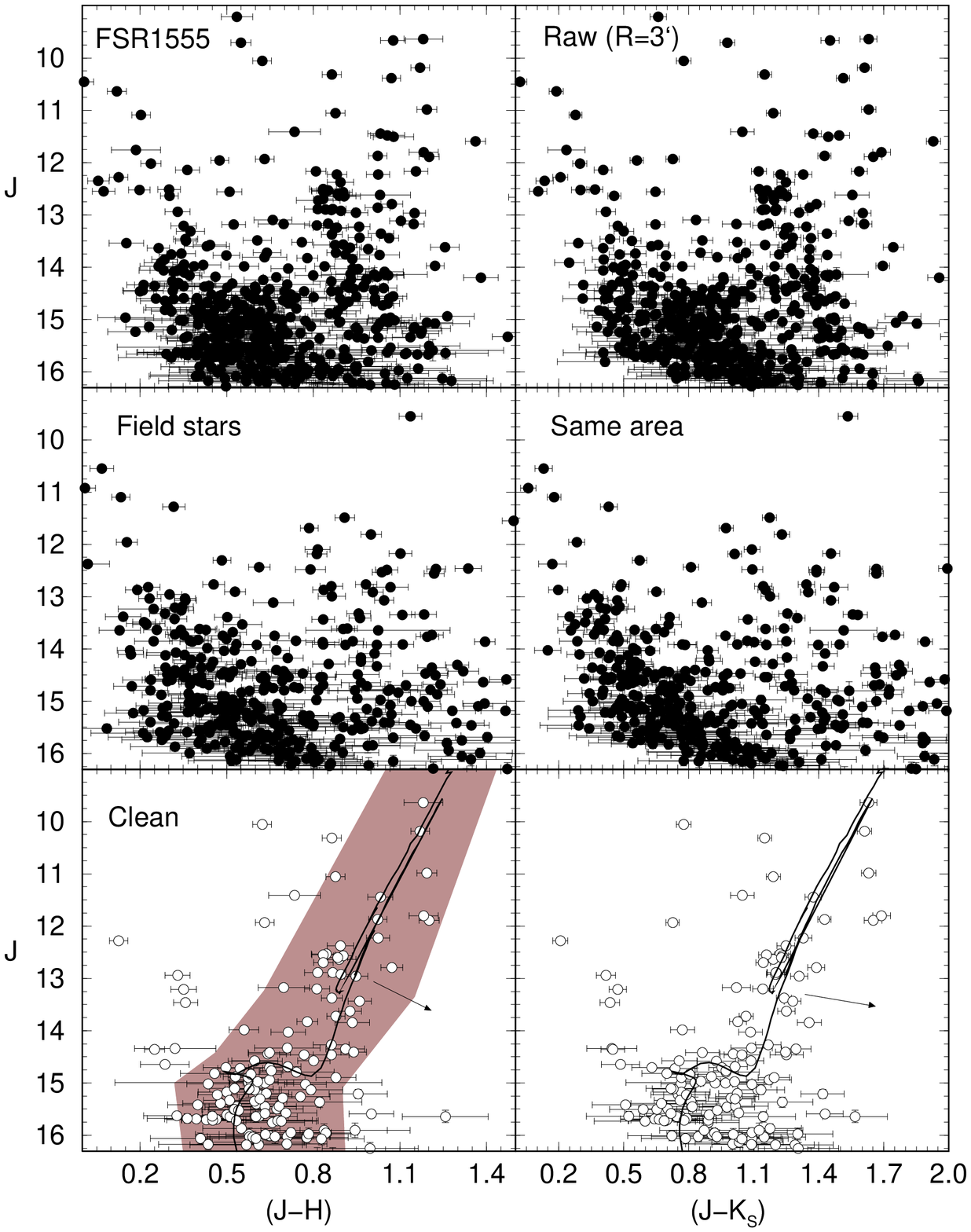}}
\caption[]{Same as Fig.~\ref{fig4} for the region $R<3\arcmin$ of FSR\,1555. The equal-area comparison 
field extraction was taken from the region $29.85\arcmin<R<30\arcmin$. The decontaminated CMDs are
best fitted with the 1.5\,Gyr Solar-metallicity Padova isochrone. Reddening vectors as in Fig.~\ref{fig4}.}
\label{fig7}
\end{figure}

CMDs displaying the $\jj\times\jh$ and $\jj\times\jk$ colours built with the raw photometry of the 
present clusters are shown in Figs.~\ref{fig4} - \ref{fig7} (top panels). For all clusters, the sampled 
region is larger than the respective core (Table~\ref{tab4}). When qualitatively compared with the CMDs 
extracted from the equal-area comparison fields (middle panels), features typical of old OCs are 
apparent. A relatively populous red clump (at $13.8\la\jj\la14.7$ and $1.3\la\jh\la1.5$) and a 
main-sequence turn off (MSTO) stand out over the field contamination of LK\,1 (Fig.~\ref{fig4}). A 
somewhat less-populous red clump ($11.3\la\jj\la12.7$, $1.2\la\jh\la1.5$) and about 3 MS mags 
($\jj\ga13.4$) are seen in LK\,10 (Fig.~\ref{fig5}). LK\,1 (Fig.~\ref{fig4}), FSR\,1521 
(Fig.~\ref{fig6}), and FSR\,1555 (Fig.~\ref{fig7}) present similar CMDs, with clear red clumps, the 
MSTO, and about 1 mag of the MS below. The red clump of FSR\,1521 occurs at $12.3\la\jj\la13.2$ and 
$0.8\la\jh\la1.0$, while for FSR\,1555 it is at $12.3\la\jj\la13.5$ and $0.7\la\jh\la1.0$. 

Finally, we show in Figs.~\ref{fig2} and \ref{fig3} the spatial distribution of the stellar surface-density 
as measured in the 2MASS photometry. We compute the surface density $\sigma$ (in units of $\rm
 stars\,arcmin^{-2}$), in a rectangular mesh with cells of dimensions $3\arcmin\times3\arcmin$. The meshes 
reach total offsets of $|\Delta\alpha|=|\Delta\delta|\approx20\arcmin$ with respect to the centre 
(Table~\ref{tab2}), in right ascension and declination; for LK\,10 we use offsets of 30\arcmin. In all
clusters, the core (Table~\ref{tab4}) is contained in the central cell. 
 
With respect to the surface-densities built with the observed (raw) photometry (top panels of Figs.~\ref{fig2} 
and \ref{fig3}), an important excess appears in the central cell, except for LK\,10 which, because of the  
contamination by disc stars, presents a rather irregular distribution. FSR\,1521 and FSR\,1555, on the
other hand, clearly detach in the central cell (Fig.~\ref{fig3}) against more uniform surrounding
fields. As shown in the bottom panels, the cluster overdensities are clearly enhanced with respect to the
surroundings in the field-star decontaminated surfaces (Sect.~\ref{Decont_CMDs}). 

\subsection{Field-star decontamination}
\label{Decont_CMDs}

As expected of low-latitude clusters (Table~\ref{tab2}), the stellar surface-density in the direction 
of the objects (Figs.~\ref{fig2} and \ref{fig3}) clearly shows that field-star contamination, 
essentially from disc stars, should be taken into account. This fact is confirmed by the qualitative 
comparison between the CMDs extracted within the cluster and in the field (Figs.~\ref{fig4}-\ref{fig7}). 
Thus, the field-star contribution should be quantified in each case to better define the intrinsic CMD 
morphology.

Field-star decontamination is a very important, yet difficult, step in the identification and
characterisation of star clusters. Several approaches have been used to this purpose (e.g. 
\citealt{Mercer05}), and most of them are based essentially on two different premises. The first works 
with spatial variations of the star-count density, but does not take into account CMD properties. In 
the latter, stars in a CMD extracted from an assumed cluster region are subtracted according to colour 
and magnitude similarity with the stars of an equal-area comparison field CMD. These methods,
together with the one we work with, are based on photometric properties only. Ideally, more robust
results on cluster membership determination would be obtained if another independent parameter, such 
as the proper motion of member and comparison field stars, is taken into account. However, for proper 
motions to be useful the target cluster should be relatively close (e.g. \citealt{AMD03}) and/or to 
have been observed in widely-apart epochs, preferentially with high resolution, such as in the case 
of the globular cluster NGC\,6397 (\citealt{Richer08}). Neither condition is satisfied for the present 
clusters, which are relatively distant (Sect.~\ref{age}) and have been observed by 2MASS in a single 
epoch.

\subsubsection{Description of the algorithm}
\label{AlgoDescr}

We work with the statistical algorithm introduced by \citet{BB07} to deal with the field-star 
contamination in CMDs. The algorithm takes into account simultaneously star-count density and 
colour/magnitude similarity between cluster and comparison field. It measures the relative number 
densities of probable field and 
cluster stars in cubic CMD cells whose axes correspond to the \jj\ magnitude and the \jh\ and \jk\ 
colours\footnote{These are the 2MASS colours that provide the maximum variance among CMD sequences for 
OCs of different ages (e.g. \citealt{TheoretIsoc}).}. The algorithm: {\em (i)} divides the full range of 
magnitude and colours covered by the CMD into a 3D grid, {\em (ii)} calculates the expected number 
density of field stars in each cell based on the number of comparison field stars with similar magnitude 
and colours as those in the cell, and {\em (iii)} subtracts the expected number of field stars from each 
cell. By construction, the algorithm is sensitive to local field-star contamination 
(\citealt{BB07}). Typical cell dimensions are $\Delta\jj=1.0$, and $\Delta\jh=\Delta\jk=0.25$, which are 
large enough to allow sufficient star-count statistics in individual cells and small enough to preserve 
the CMD morphology. The comparison fields are located within $R=20\arcmin - 
40\arcmin$ (LK\,1), $R=30\arcmin-60\arcmin$ (LK\,10), and $R=15\arcmin-30\arcmin$
(FSR\,1521 and FSR\,1555). 
The inner boundary of the comparison field lies beyond the probable tidal radius 
(Sect.~\ref{struc}), which minimises the probability of oversubtraction of cluster stars. We emphasise 
that the equal-area field extractions (middle panels of Figs.~\ref{fig4} - \ref{fig7}) should be taken
only for qualitative comparisons. The decontamination is based on the large surrounding area as described 
above. Further details on the algorithm are given in \citet{BB07}.

\begin{table*}
\caption[]{Field-star decontamination statistics}
\label{tab3}
\renewcommand{\tabcolsep}{2.95mm}
\renewcommand{\arraystretch}{1.2}
\begin{tabular}{cccccccccccccc}
\hline\hline
$\Delta\jj$&&\multicolumn{5}{c}{LK\,1 ($R<2\arcmin$) - $f_{sub}=97.5\%$}&
&\multicolumn{5}{c}{LK\,10 ($R<5\arcmin$) - $f_{sub}=91.5\%$ }\\
\cline{3-7}\cline{9-13}
 &&\no&\nc&\ns&\sFS&\fsU& &\no&\nc&\ns&\sFS&\fsU\\
(mag)&&(stars)&(stars)&&(stars)&&&(stars)&(stars)&&(stars)\\
\cline{1-13}
  8--9&&   ---      &---&---&---&---&    &$1\pm1.0$   &1 &1.0&0.21&0.33    \\
 9--10&&   ---      &---&---&---&---&    &$1\pm1.0$   &0 &0.0&0.61&0.38    \\
10--11&&$1\pm1.0$   &1 &1.0&0.18&0.18&   &$8\pm2.8$   &6 &2.1&0.70&0.19 \\
11--12&&$2\pm1.4$   &1 &0.7&0.23&0.09& &$24\pm4.9$  &15&3.1&2.46&0.28 \\
12--13&&$6\pm2.4$   &1&0.4&0.35&0.06&  &$29\pm5.9$  &12&2.2&5.66&0.28  \\
13--14&&$21\pm4.6$  &15&3.3&1.38&0.11&  &$67\pm8.2$  &32&3.9&11.94&0.29  \\
14--15&&$58\pm7.6$  &32&4.2&2.65&0.10&  &$134\pm11.6$&67&5.8&25.83&0.32  \\
15--16&&$87\pm9.3$  &34&3.6&3.76&0.08&  &$224\pm15.0$&79&5.3&56.14&0.35  \\
16--16.5&&$57\pm7.5$&29&3.8&6.93&0.22&  &$261\pm16.2$&74&4.6&32.87&0.16  \\
\cline{1-1}\cline{3-7}\cline{9-13}
8--16.5&&$232\pm15.2$  &113&7.2&10.7&0.09&&$747\pm27.3$&286&8.9&15.6&0.21 \\
\hline
\hline
$\Delta\jj$&&\multicolumn{5}{c}{FSR\,1521 ($R<4\arcmin$) - $f_{sub}=97.9\%$}&
&\multicolumn{5}{c}{FSR\,1555 ($R<3\arcmin$) - $f_{sub}=96.6\%$}\\
\cline{3-7}\cline{9-13}
 &&\no&\nc&\ns&\sFS&\fsU& &\no&\nc&\ns&\sFS&\fsU\\
(mag)&&(stars)&(stars)&&(stars)&&&(stars)&(stars)&&(stars)\\
\cline{1-13}
 8--9&&$2\pm1.3$    &1  &0.7&0.87&0.87  &&   ---      &---&---&---&---\\
 9--10&&  ---       &---&---&--- &---  &&$1\pm1.0$   &0 &0.0&0.47&0.33\\
10--11&&$2\pm1.4$   &1  &0.7&1.09&0.21 &&$2\pm1.4$   &2 &1.4&0.91&0.12\\
11--12&&$18\pm4.2$  &6  &1.4&1.82&0.14 &&$9\pm3.0$   &6 &2.0&1.08&0.61\\
12--13&&$41\pm6.4$  &21 &3.3&1.81&0.07 &&$20\pm4.5$  &11&2.5&1.50&1.34\\
13--14&&$59\pm7.7$  &13 &1.7&3.04&0.06 &&$27\pm5.2$  &12&2.3&2.33&1.94\\
14--15&&$155\pm12.4$&42 &3.4&12.32&0.11 &&$106\pm10.3$&62&6.0&3.77&4.52\\
15--16&&$270\pm16.4$&64 &3.9&10.44&0.05 &&$121\pm11.0$&57&5.2&5.61&10.13\\
\cline{1-1}\cline{3-7}\cline{9-13}
8--16&&$547\pm23.4$&146&6.1&15.3&0.03 &&$286\pm16.9$&150&8.2&16.3&0.08\\
\hline
\end{tabular}
\begin{list}{Table Notes.}
\item For each magnitude bin ($\Delta\jj$), we give the number of observed stars 
(\no) within the spatial region sampled in the CMDs shown in the top panels of Figs.~\ref{fig4} 
and \ref{fig5}, the respective number of probable member stars (\nc) computed by the 
decontamination algorithm, the \ns\ parameter, the $1\,\sigma$ Poisson fluctuation (\sFS) around 
the mean, with respect to the star counts measured in the 8 sectors of the comparison field, and the 
field-star uniformity parameter. The statistical significance of \nc\ is reflected in its ratio with 
the $1\sigma$ Poisson fluctuation of \no\ (\ns) and with \sFS. The bottom line corresponds to the  
full magnitude range. The subtraction efficiency ($f_{sub}$) is also given.
\end{list}
\end{table*}

As discussed in \citet{BB07}, differential reddening between cluster and field stars may be critical 
for the decontamination algorithm. Important gradients would require large cell sizes or, in extreme cases, 
preclude application of the algorithm altogether. Basically, it would be required colour differences of
$|\Delta\jh|\ga\rm cell~size$ (0.25, in the present work) between cluster and comparison field for the 
differential reddening to affect the subtraction in a given cell. However, the cluster and comparison field 
CMDs (Figs.~\ref{fig4}-\ref{fig7}) indicate that differential reddening is not important for the present 
sample.

\subsubsection{Decontamination statistics}
\label{DecStat}

Additional statistical analysis is required because of the relatively high reddening values 
(Table~\ref{tab2}) affecting the clusters. In Table~\ref{tab3} we present the full statistics of the 
decontamination, discriminated by magnitude bins. Statistically relevant parameters are: {\em (i)} \ns\ 
which, for a given magnitude bin, corresponds to the ratio of the decontaminated number of stars to the 
$1\sigma$ Poisson fluctuation of the number of observed stars, {\em (ii)} \sFS, which is related to 
the probability that the decontaminated stars result from the normal star count fluctuation in the comparison 
field and, {\em (iii)} \fsU, which measures the star-count uniformity of the comparison field. Properties of 
\ns, \sFS, and \fsU, measured in OCs and field fluctuations are discussed in \citet{ProbFSR}. Table~\ref{tab3}
also provides integrated values of the above parameters, which correspond to the full magnitude range spanned
by the CMD of each OC. The spatial regions are those sampled by the CMDs shown in the top panels 
of Figs.~\ref{fig4}-\ref{fig7}.

CMDs of star clusters should have integrated \ns\ values significantly higher than 1 (\citealt{ProbFSR}),
a condition that is met for the present objects ($\ns=6.1-8.9$). As a further test of the statistical significance 
of the above results, we investigate star count properties of the field stars. First, the comparison field 
is divided into 8 sectors around the cluster centre. Next, we compute the parameter \sFS, which is the 
$1\,\sigma$ Poisson fluctuation around the mean of the star counts measured in the 8 sectors (corrected 
for the different areas of the sectors and cluster extraction). In a spatially uniform comparison field, 
\sFS\ is expected to be very small. Thus, OCs should have the probable number of member 
stars (\nc) higher than $\sim3\,\sFS$, to minimise the probability that \nc\ arises from fluctuations of a 
non-uniform comparison field. This condition is fully satisfied by the present clusters, reaching the level 
$\nc\sim(5-10)\,\sFS$. We also provide in Table~\ref{tab3} the parameter \fsU. For a given magnitude bin we 
first compute the average number of stars over all sectors $\langle N\rangle$ and the corresponding 
$1\sigma$ fluctuation $\sigma_{\langle N\rangle}$; thus, \fsU\ is defined as $\fsU=\sigma_{\langle
N\rangle}/\langle N\rangle$. Non uniformities such as heavy differential reddening should result in high 
values of \fsU. On the other hand, \fsU\ is low ($\la0.09$), except for LK\,10 ($\fsU=0.21$), which reflects a rather
irregular field (Fig.~\ref{fig2}). Finally, we note that the number of decontaminated stars in most magnitude 
bins is larger than what could be expected from field-star fluctuations ($\nc\ga3\sFS$).

Since we usually work with comparison fields larger than the possible-cluster extractions, the correction
for the different spatial areas between field and cluster is expected to produce a fractional number
of probable field stars ($n_{fs}^{cell}$) in some cells. Before the cell-by-cell subtraction, the
fractional numbers are rounded off to the nearest integer, but limited to the number of observed stars
in each cell $n_{sub}^{cell}=NI(n_{fs}^{cell})\leq n_{obs}^{cell}$, where NI represents rounding off to
the nearest integer). The global effect is quantified by means of the difference between the expected number
of field stars in each cell ($n_{fs}^{cell}$) and the actual number of subtracted stars ($n_{sub}^{cell}$).
Summed over all cells, this quantity provides an estimate of the total subtraction efficiency of the
process, \[ f_{sub}=100\times\sum_{cell}n_{sub}^{cell}/\sum_{cell}n_{fs}^{cell}~~~~~(\%).\] Ideally, the best
results would be obtained for an efficiency $f_{sub}\approx100\%$. With the assumed grid settings for the
decontamination of the present clusters, the subtraction efficiencies turned out to be higher than 90\%.

\subsubsection{Decontaminated CMDs and surface densities}
\label{DecOut}

As an indicator of the algorithm efficiency we can take the decontaminated stellar surface-density 
distributions (bottom panels of Figs.~\ref{fig2} and \ref{fig3}). The central excesses have been 
significantly enhanced with respect to the raw photometry (top panels), while the residual 
surface-density around the centre has been reduced to a minimum level. By design, the decontamination
depends essentially on the colour-magnitude distribution of stars located in different spatial regions.
The fact that the decontaminated surface-density presents a conspicuous excess only at the assumed 
cluster position implies significant differences among this region and the comparison field, both in 
terms of colour-magnitude and number of stars within the corresponding colour-magnitude bins. This 
meets the expectations of star clusters, which can be characterised by a 
single-stellar population, projected against a Galactic stellar field. 

The decontaminated CMDs are shown in the bottom panels of Figs.~\ref{fig4} - \ref{fig7}. As expected, 
essentially all of the disc contamination is removed, leaving stellar sequences typical of reddened 
old OCs, with well-developed red clumps and different extents of the MS.

As a caveat, we cannot exclude the possibility of differential reddening to account for part
of the observed spread in the CMD distribution of stars. To examine this issue we show, in the 
bottom panels of Figs.~\ref{fig4}-\ref{fig7}, reddening vectors computed with the 2MASS ratios 
(Sect.~\ref{2mass}) for a standard visual absorption $\aV=2$. Given the absorptions derived for 
the clusters (Table~\ref{tab2}), the standard value, which represents from $\approx25\%$ to 
$\approx50\%$ of the total \aV, can be taken as an upper limit to the differential reddening.
Together with the decontaminated CMDs, this experiment suggests that differential reddening in 
all cases is not large, because the giant clumps are rather tight, while the MS spread appears 
to be dominated by photometric errors. The typical dependence of the 2MASS photometric errors on 
magnitude, for objects projected along different directions, is discussed in \citet{OldOCs}. 

We conclude that the qualitative and quantitative expectations of the decontamination algorithm have 
been satisfied by the output. In all cases, the decontaminated photometry presents a conspicuous excess, 
with respect to the surroundings, in the surface-density distribution (Figs.~\ref{fig2} and \ref{fig3}). 
In addition, field-decontaminated CMDs extracted from the spatial regions where the excesses occur
(Figs.~\ref{fig4}-\ref{fig7}), present statistically significant (Table~\ref{tab3}) cluster CMDs.

\section{Cluster age, reddening, and distance}
\label{age}

The field-decontaminated CMD morphologies derived in Sect.~\ref{2mass} can be used to compute the 
cluster fundamental parameters.{\bf We work with Padova isochrones (\citealt{Girardi2002}) computed 
with the 2MASS \jj, \hh, and \ks\ filters\footnote{\em http://stev.oapd.inaf.it/cgi-bin/cmd - 
Bolometric and colour corrections were computed for a set of isochrones using the 2MASS filter 
responses. The isochrones were subsequently provided in the Vega Mag system.}. } The updated isochrones 
are very similar to the Johnson-Kron-Cousins ones (e.g. \citealt{BesBret88}), with differences of at 
most 0.01 in \jh\ (\citealt{TheoretIsoc}). Distinctive features of the updated isochrone set are 
centred mostly on the greatly-improved treatment of the thermally-pulsing asymptotic giant branch 
(TP-AGB) phase. According to \citet{Marigo08}, the updated isochrones are intended to preserve the 
several peculiarities present in the TP-AGB tracks, namely, the cool tails of C-type stars due to the 
use of proper molecular opacities as convective dredge-up occurs along the TP-AGB, the bell-shaped 
sequences in the HR diagram for stars with hot-bottom burning, the changes of pulsation mode between 
fundamental and first overtone, the sudden changes of mean mass-loss rates as the surface chemistry 
changes from M- to C-type, etc. Because it is important to derive, or at least set constraints on 
astrophysical parameters of old star clusters, we adopt as working strategy the search for solutions 
within a range of ages and metallicities. As discussed by, e.g. \citet{Friel95}, OC metallicities in 
general range from Solar ($[Fe/H]=0$, or $Z=0.019$) to sub-Solar ($[Fe/H]=-0.5$, $Z=0.006$, or 
$\approx1/3$ Solar) values. According to the metallicity gradient (Fig.~7 in \citealt{Friel95}), the 
present OCs fall around the locus occupied by OCs with metallicity $[Fe/H]\approx-0.1$. Thus we base 
the following analysis on this most probable value. To compute Galactocentric distances, we adopt
$\rs=7.2\pm0.3$\,kpc (\citealt{GCProp}) as the Sun's distance to the Galactic centre. This value
was derived by means of the GC spatial distribution\footnote{Other recent studies gave similar results, 
e.g. $\rs=7.2\pm0.9$\,kpc (\citealt{Eisen03}), $\rs=7.62\pm0.32$\,kpc (\citealt{Eisen05}) and 
$\rs=7.52\pm0.10$\,kpc (\citealt{Nishiyama06}), with different approaches.}. 

Historically, different approaches have been used to extract astrophysical parameters from
CMDs by means of isochrone fits. The simplest ones are based on a direct comparison of a set
of isochrones with the CMD morphology, while the more sophisticated include photometric
uncertainties, binarism, and variations on metallicity. Most of these methods are summarised in 
\citet{NJ06}, in which a maximum-likelihood CMD fit method is described. 

For simplicity, in the present cases fits are made {\em by eye}, with the tight giant clumps as
the strongest constraint. We also require that, because of the potential presence of binaries, 
the adopted isochrone should be shifted somewhat to the left of the MS fiducial line, i.e. a 
median line that takes into account the MS spread, including the photometric uncertainties as 
well (e.g. \citealt{N188}, and references therein). In the following Section we discuss each 
cluster individually. 

\subsection{FSR\,1521}
\label{FSR1521}

The decontaminated CMD morphology of FSR\,1521 (Fig.~\ref{fig6}) shows a concentration of 
stars at $\jj\approx15$, which indicates a relatively populous MSTO of an old cluster detected near
the 2MASS photometric limit. Besides, a distinctive red clump shows up at $\jj\approx12.5$ and 
$\jh\approx0.9$. Taken together, these stellar sequences characterise a distant old OC. 

\begin{figure}
\resizebox{\hsize}{!}{\includegraphics{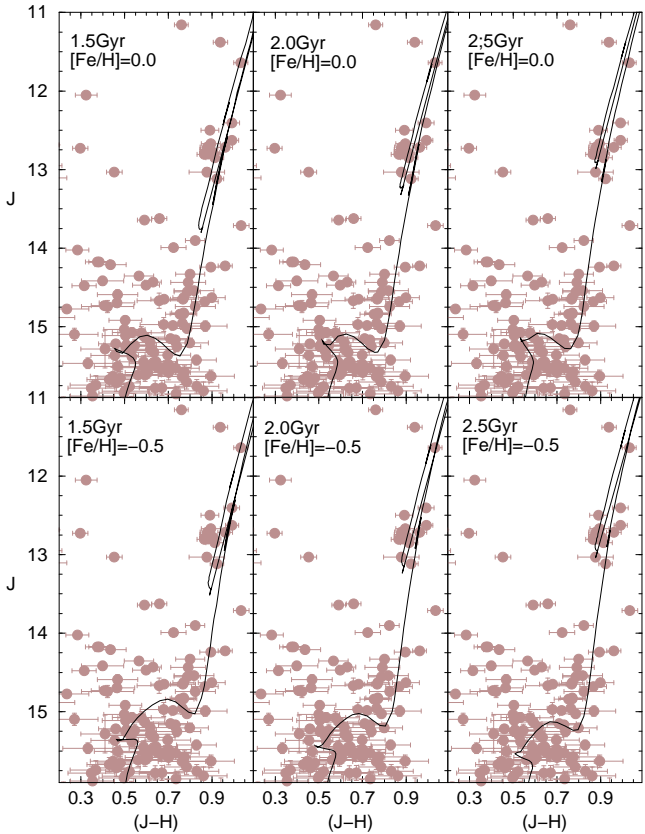}}
\caption[]{Possible solutions for the age and metallicity of FSR\,1521. Padova isochrones with
the ages 1.5 (left panels), 2 (middle), and 2.5\,Gyr (right) are used. Except for the Solar
metallicity 1.5\,Gyr isochrone, the remaining ones provide acceptable fits, either with 
Solar-metallicity (top panels) or the $1/3$ Solar-metallicity (bottom) solutions.}
\label{fig8}
\end{figure}

Allowing as well for the photometric uncertainties, acceptable fits to the decontaminated 
CMD morphology are obtained with the 2\,Gyr isochrone, with an uncertainty of $\pm0.5$\,Gyr. This can 
be seen in Fig.~\ref{fig8}, in which we compare solutions with the ages 1.5, 2, and 2.5\,Gyr, both for 
Solar and sub-Solar ($[Fe/H]=-0.5$) metallicities. Ages younger than about 1.5\,Gyr would produce a 
significantly poorer description (much fainter than observed) of the red clump, while those older 
than about 2.5\,Gyr, would be 
prohibitively shifted to the right of the MSTO. In any case, both metallicity ranges provide similar 
fits to the decontaminated morphology. Since the remaining OCs have ages (Table~\ref{tab2}) and CMDs 
with photometric uncertainties similar to those of FSR\,1521, the above test shows that any
metallicity within $1/3\la Z/Z_\odot\la1$ would produce acceptable solutions. Thus, for simplicity 
and because the present OCs are relatively near the Solar circle, we adopt the Solar-metallicity isochrones 
as more probable solutions.

We caution that, because of the 2MASS photometric uncertainties for the lower sequences,
a more sophisticated approach for isochrone fitting might produce an overinterpretation. For
this reason, we apply the above method for all objects.

With the adopted solution, the fundamental parameters of FSR\,1521 are the near-infrared reddening 
$\ejh=0.34\pm0.05$, which corresponds to $\ebv=1.1\pm0.2$, or
$A_V=3.4\pm0.5$, the observed and absolute distance moduli $\mMJ=14.2\pm0.3$ and $\mMo=13.26\pm0.33$,
respectively, and the distance from the Sun $\ds=4.5\pm0.7$\,kpc. Thus, for $\rs=7.2$\,kpc, the 
Galactocentric distance of FSR\,1521 is $\dgc=7.8\pm0.4$\,kpc, which puts it $\approx0.6$\,kpc outside 
the Solar circle. This solution is shown in Fig.~\ref{fig6} (bottom panels). 

We call attention that the best fit takes into account strong constraints provided by the 
tight distribution of the clump giants, which preserves accuracy especially for the reddening 
and distance. Thus, the resulting uncertainties are adopted as the most probable errors, which
are subsequently propagated to the different sets of parameters (Tables~\ref{tab2} and 
\ref{tab4}). However, given the uncertainties in the faint-photometric end, age is less 
constrained.


\subsection{LK\,1}
\label{LK1}

The {\em best-fit} to the CMD of LK\,1 was obtained with the 1\,Gyr isochrone, $\ejh=0.90\pm0.03$ and
$\mMJ=15.5\pm0.2$. Taking into account fit uncertainties we derive the age $1.0\pm0.2$\,Gyr, 
$\ebv=2.9\pm0.1$, $A_V=8.9\pm0.3$, $\mMo=13.02\pm0.22$, $\ds=4.0\pm0.4$\,kpc, and $\dgc=7.3\pm0.2$\,kpc. 
LK\,1 is $\approx0.1$\,kpc outside the  Solar circle. This solution is shown in the bottom panels of
Fig.~\ref{fig4}. It represents well the MSTO, red clump and giant branch, in both colours. The few stars
bluer than the MSTO might be blue stragglers or un-subtracted field stars. Our derived distance 
modulus is about 2 mags fainter than that adopted by \citet{LK2002}.

With 1-2\,kpc as the distance to the Cygnus Association (e.g. \citealt{LK2002}; \citealt{BBD2003}), 
LK\,1, at $\ds\approx4$\,kpc, is a distant background old open cluster in the line of sight of 
that association.

\subsection{LK\,10}
\label{LK10}

The decontaminated CMD morphology of LK\,10 (Fig.~\ref{fig5}) is somewhat more constrained than that 
of LK\,1. Besides a tight red clump, it features about a 3 mag MS extent. Fundamental parameters of 
LK\,10 are an age $1.0\pm0.1$\,Gyr, $\ejh=0.86\pm0.02$, which corresponds to $\ebv=2.8\pm0.1$ and 
$A_V=8.5\pm0.3$, $\mMJ=13.1\pm0.1$, $\mMo=10.7\pm0.11$, $\ds=1.4\pm0.1$\,kpc, and $\dgc=7.1\pm0.1$\,kpc. 
LK\,10 lies $\approx0.1$\,kpc inside the Solar circle. Our derived distance modulus agrees with that 
adopted by \citet{LK2002}.

With $\ds\sim1.4$\,kpc, LK\,10 might be spatially coincident with the Cygnus Association. However, 
they must not share a common origin, since LK\,10 is much older than the estimated age of the 
Cygnus Association, 1-4\,Myr (e.g. \citealt{Massey95}). Despite the different distances, LK\,1 and 
LK\,10 are affected essentially by the same reddening value.

\subsection{FSR\,1555}
\label{FSR1555}

With the MSTO and red clump detected in the decontaminated CMD of FSR\,1555 (Fig.~\ref{fig7}), we derive 
an age $1.5\pm0.5$\,Gyr, $\ejh=0.37\pm0.02$, which corresponds to $\ebv=1.2\pm0.1$ and $A_V=3.7\pm0.3$,
$\mMJ=14.1\pm0.1$, $\mMo=13.08\pm0.11$, $\ds=4.1\pm0.2$\,kpc, and $\dgc=7.1\pm0.1$\,kpc. FSR\,1555 lies
$\approx0.1$\,kpc inside the Solar circle. 

\section{Cluster structure}
\label{struc}

\begin{figure}
\resizebox{\hsize}{!}{\includegraphics{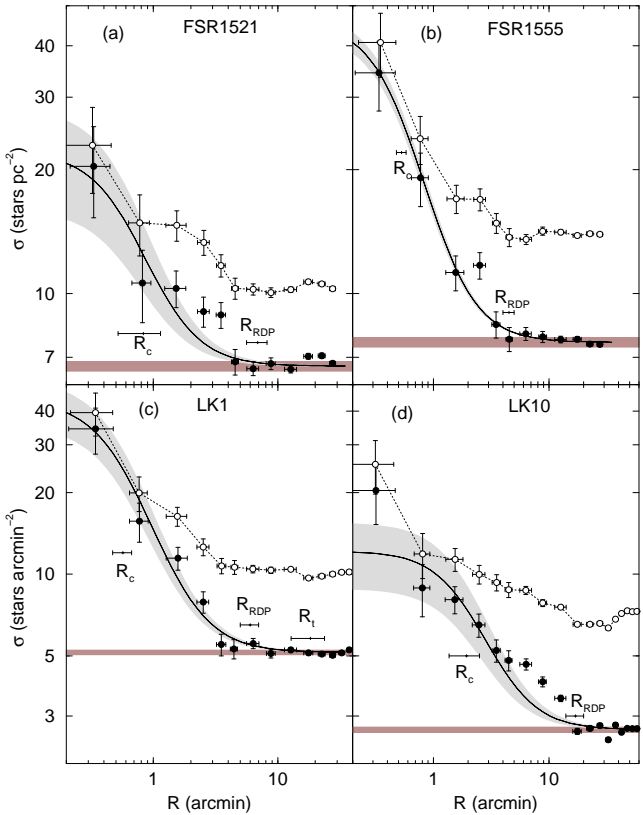}}
\caption[]{Stellar RDPs built with colour-magnitude filtered (filled circles) and raw (empty) 
photometry. Solid line: best-fit King-like profile. Horizontal shaded polygon: background stellar 
level. Shaded regions: $1\sigma$ King fit uncertainty. The core (\rc) and cluster (\rl) radii are 
indicated in all cases, while for LK\,1, the tidal radius (\rt) is also shown. Note the central 
density excess in LK\,10 profile.}
\label{fig9}
\end{figure}

Structural parameters are derived by means of the projected radial density profiles (RDP) built
with the stellar number density around the cluster centre. Usually, star clusters have RDPs that 
follow a well-defined analytical profile. Among these are the empirical, single mass, modified 
isothermal sphere of \citet{King66}, the modified isothermal sphere of \citet{Wilson75}, which
assumes a pre-defined stellar distribution function (and produces more extended envelopes than 
King 1966), and the power law with a core of \citet{EFF87}. These functions are characterised 
by different parameters that are related to cluster structure. However, considering the error 
bars of the present RDPs (Fig.~\ref{fig9}), we adopt the analytical 
function $\sigma(R)=\sigma_{bg}+\sigma_0/(1+(R/R_c)^2)$, where $\sigma_{bg}$ is the residual 
background density, $\sigma_0$ is the central density of stars, and \rc\ is the core radius. This 
function is similar to that introduced by \cite{King1962} to describe the surface brightness profiles 
in the central parts of globular clusters. As discussed in \citet{StrucPar}, RDPs built with
depth-limited photometry produce structural radii comparable to the intrinsic (i.e. derived
with deep photometry) ones.


\begin{table*}
\caption[]{Derived cluster structural parameters}
\label{tab4}
\renewcommand{\tabcolsep}{0.95mm}
\renewcommand{\arraystretch}{1.25}
\begin{tabular}{lccccccccccccc}
\hline\hline
Cluster&$\sigma_{bg}$&$\sigma_0$&\rc&\rl&\rt&$\delta_c$&$1\arcmin$&$\sigma_{bg}$&$\sigma_0$&\rc&\rl&\rt\\
       &$\rm(*\,\arcmin^{-2})$&$\rm(*\,\arcmin^{-2})$&(\arcmin)&(\arcmin)&(\arcmin)& &(pc)&
$\rm(*\,pc^{-2})$&$\rm(*\,pc^{-2})$&(pc)&(pc)&(pc)\\
(1)&(2)&(3)&(4)&(5)&(6)&(7)&(8)&(9)&(10)&(11)&(12)&(13)\\
\hline
LK\,1&$5.15\pm0.03$&$38.5\pm8.3$&$0.57\pm0.10$&$6.0\pm1.0$&$18\pm5$&$8.5\pm1.6$&1.163&$3.8\pm0.1$&$28.5\pm6.1$&
   $0.66\pm0.12$&$7.0\pm1.2$&$21\pm6$\\
   
LK\,10&$2.67\pm0.02$&$9.5\pm3.3$&$1.94\pm0.57$&$16.0\pm2.0$&---&$4.6\pm1.2$&0.444&$16.3\pm0.1$&$58.1\pm17.1$&
   $0.78\pm0.23$&$6.5\pm0.8$&---\\
   
FSR\,1521&$6.66\pm0.06$&$15.4\pm7.1$&$0.64\pm0.24$&$5.2\pm1.0$&---&$3.3\pm1.0$&1.302&$3.9\pm0.1$&$9.1\pm4.2$&
   $0.83\pm0.31$&$6.9\pm1.3$&---\\
   
FSR\,1555&$7.62\pm0.03$&$37.8\pm2.6$&$0.53\pm0.05$&$4.5\pm0.5$&---&$6.0\pm0.3$&1.197&$5.3\pm0.1$&$26.4\pm1.8$&
   $0.63\pm0.06$&$5.4\pm0.6$&---\\
\hline
\end{tabular}
\begin{list}{Table Notes.}
\item Core (\rc), cluster (\rl), and tidal (\rt) radii are given in angular and absolute units. 
Col.~7: cluster/background density contrast parameter ($\delta_c=1+\sigma_0/\sigma_{bg}$), 
measured in the colour-magnitude filtered RDPs. Col.~8: arcmin to parsec scale. 
\end{list}
\end{table*}

To minimise noise in the RDPs, we first apply a colour-magnitude filter to the photometry, which
excludes stars with colours unlike those of the cluster sequence. Colour-magnitude 
filters are wide enough to include cluster MS and evolved star colour distributions, as well as the 
$1\sigma$ photometric uncertainties\footnote{Colour-magnitude filter widths should also account for 
formation or dynamical evolution-related effects, such as enhanced fractions of binaries (and other 
multiple systems) towards the central parts of clusters, since such systems tend to widen the MS 
(e.g. \citealt{HT98}; \citealt{Kerber02}; \citealt{BB07}; \citealt{N188}).}. The colour-magnitude 
filters for the present OCs are shown in the bottom-left panels of Figs.~\ref{fig4}-\ref{fig7}. 
However, residual field stars with colours similar to those of the cluster are expected to remain 
inside the colour-magnitude filter. They affect the intrinsic stellar RDP in a way that 
depends on the relative densities of field and cluster stars. The contribution of the residual 
contamination to the observed RDP is statistically subtracted by means of the field. As a 
result, the use of colour-magnitude filters enhances the contrast of the RDP 
with respect to the background, especially in crowded fields (e.g. \citealt{BB07}).

Oversampling near the centre and undersampling at large radii are avoided by using rings
of increasing width with distance from the cluster centre. A typical set of ring widths is
$\Delta\,R=0.5,\ 1,\ 2.5,\ {\rm and}\ 5\arcmin$, respectively for $0\arcmin\le R<1\arcmin$,
$1\arcmin\le R<4\arcmin$, $4\arcmin\le R<10\arcmin$, and $R\ge10\arcmin$. The number and
width of the rings can be set to produce RDPs with adequate spatial resolution and small
$1\sigma$ Poisson errors. The residual background level of each RDP corresponds to the average
number of colour-magnitude filtered stars measured in the field. The $R$ coordinate (and
uncertainty) of each ring corresponds to the average position and standard deviation of the
stars inside the ring.

The colour-magnitude filtered RDPs of the present clusters are shown in Fig.~\ref{fig9}, where we
also show the profiles produced with the observed (raw) photometry. As expected, 
minimisation of the number of non-cluster stars by the colour-magnitude filter resulted in RDPs with 
higher contrast with respect to the background. Fits of the King-like profile were performed with a 
non-linear least-squares fit routine that uses errors as weights. To minimise degrees of freedom, 
$\sigma_0$ and \rc\ were derived from the RDP fit, while $\sigma_{bg}$ is measured in the field. 
The best-fit solutions are shown in Fig.~\ref{fig9}, and the fit parameters are given in Table~\ref{tab4}.
For absolute comparison with other clusters, Table~\ref{tab4} also presents parameters in absolute
units, based on the cluster distances (Sect.~\ref{age}). Because of 
the 2MASS photometric limit, which for the present clusters corresponds to a cutoff for stars brighter 
than $\jj\approx16.5$, $\sigma_0$ should be taken as a lower limit.

Within uncertainties, the adopted King-like function describes well the colour-magnitude filtered RDPs 
along the full radius range, especially for LK\,1, FSR\,1521, and FSR\,1555. The exception is LK\,10, 
which shows a marked excess in the central region.
This central cusp in LK\,10 suggests a post-core collapse phase in this $\sim1$\,Gyr OC, as detected 
in part of the globular clusters (e.g. \citealt{TKD95}). Such a central cusp has been observed in the
RDP of other OCs, such as NGC\,3960 (\citealt{N3960}).

We also estimate the cluster radius (\rl) by visually comparing the RDP level (and fluctuations) with 
the background. It corresponds to the distance from the cluster centre where RDP and background are 
statistically indistinguishable (e.g. \citealt{DetAnalOCs}, and references therein). Thus, most of the 
cluster stars are contained within $\rl$, which should not be mistaken for the tidal radius. Tidal radii 
are derived from, e.g. the 3-parameter King-profile fit to RDPs (see below), which requires large 
surrounding fields and adequate Poisson errors. For instance, in populous and relatively high Galactic 
latitude OCs such as M\,67, NGC\,188, and NGC\,2477, cluster radii are a factor $\sim0.5 - 0.7$ of the 
respective tidal radii (\citealt{DetAnalOCs}). The factor is somewhat lower for low-Galactic latitude 
objects, because of the enhanced background. The cluster radii of the present objects are given in 
cols.~5 (angular scale) and 12 (absolute scale) of Table~\ref{tab4}. 

Table~\ref{tab4} (col.~7) provides the density contrast parameter $\delta_c=1+\sigma_0/\sigma_{bg}$,
which, for the present clusters is relatively high ($3.3<\delta_c<8.5$). Since $\delta_c$ is measured 
in colour-magnitude-filtered (lower noise) RDPs, it is usually higher than the visual contrast produced 
by images (e.g. Fig.~\ref{fig1}).

Alternatively, we tried to fit the RDPs with the 3-parameter function (based on \citealt{King1962})
$$\sigma(R)=\sigma_0\left[\frac{1}{\sqrt{1+(R/R_c)^2}}-\frac{1}{\sqrt{1+(R_t/R_c)^2}}\right]^2,$$
which includes the tidal radius (\rt). However, convergence occurred only for LK\,1, with 
$\rt=18.3\arcmin\pm5.5\arcmin$. With such radii, the concentration parameter of LK\,1 is 
$c=\log(\rt/\rc)\approx1.5$ which, compared to the Galactic (non-core collapse) globular clusters, 
puts it around the median value (e.g. \citealt{Pap11GCs}).
Qualitatively, the 3-parameter fit to the RDP of LK\,1 is indistinguishable to that shown in 
Fig.~\ref{fig9}, within the uncertainties. Thus, we simply show in panel (c) the position of the 
tidal radius.

Interestingly, the cluster radii of LK\,1 and LK\,10 given by \citet{LK2002} are about half those 
derived in this work. This probably occurs because we work with colour-magnitude-filtered RDPs,
which enhances the cluster/background contrast and probes larger cluster extensions. With 
respect to FSR\,1521 and FSR\,1555, our values for the core radius are about $1/3$ of those given by
\citet{FSRcat}, while our \rl\ is consistent with their \rt, given the above relation between
both radii. 

Compared to the distribution of core radius derived for a sample of relatively nearby OCs by 
\citet{Piskunov07}, the present OCs, especially FSR\,1521, occupy the small-\rc\ tail. Besides,
for a relation between tidal and cluster radius as $\sim2\times\rl$, LK\,10, FSR\,1521, and
FSR\,155 are located around the median value, while LK\,1 occupies the large-\rt\ tail.

\section{Mass estimates}
\label{MF}

Since we detect $\approx3$ mags of the MS of LK\,10, we can build its mass function
$\left(\phi(m)=\frac{dN}{dm}\right)$ and compute the mass stored in stars. We work with colour-magnitude filtered 
photometry, the 3 2MASS bands separately, and the mass-luminosity relation obtained from the corresponding 
Padova isochrone and distance from the Sun (Sect.~\ref{age}). Further details on MF construction are given 
in \citet{FaintOCs}. The effective MS stellar mass range is $1.15\leq m(\ms)\leq1.80$.

The MF of LK\,10 is well represented by the function $\phi(m)\propto m^{-(1+\chi)}$, with
the slope $\chi=2.4\pm0.4$. Within uncertainties, this slope agrees with $\chi=1.98\pm0.25$ 
derived by \citet{LK2002}. Both values are steeper than the $\chi=1.35$ of 
\citet{Salpeter55} initial mass function (IMF).

The number of observed MS and evolved stars in LK\,10 (for $R\le\rl$) is derived by counting the 
stars (in the background-subtracted colour-magnitude filtered photometry) that are present in the
mag ranges, $13.4<\jj<16.4$ for the MS and $\jj<13.4$ for the evolved stars. There are 
$n_{MS}=959\pm20$ and $n_{evol}=25\pm8$, MS and evolved stars, respectively; the corresponding
mass values are $m_{MS}=1311\pm26\,\ms$ and $m_{evol}=46\pm16\,\ms$. The evolved star mass corresponds 
to $n_{evol}$ multiplied by the stellar mass at the TO, $m_{TO}=1.8\,\ms$. Thus, the observed stellar 
mass of LK\,10 is $m_{obs}\approx1360\,\ms$, which agrees with the equivalent value estimated by 
\citet{LK2002}.
 
Finally, we estimate the total stellar mass by extrapolating the observed MF down to the H-burning 
mass limit ($0.08\,\ms$). We follow the universal IMF of \citet{Kroupa2001}, which assumes increasing 
flattening towards low-mass stars. This IMF is described by the slopes $\chi=0.3\pm0.5$ for the range 
$0.08\leq m(\ms)\leq0.5$ and $\chi=1.3\pm0.3$ for $0.5\leq m(\ms)\leq1.0$. We obtain 
$m_{extr}=4420\pm2000\,\ms$. Thus, the stellar mass of LK\,10 can be put in the range 
$1360-4420\,\ms$. Again, within uncertainties, the upper value is consistent with that 
estimated by \citet{LK2002}.

Because of the limited MS range, only estimates of the observed cluster mass are made for
the remaining objects, by means of the age solutions given in Sect.~\ref{age}. In all cases
we consider the region within $R\le\rl$ (Table~\ref{tab4}). For LK\,1 we derive 
$m_{MS}=108\pm9\,\ms$ and $m_{evol}=270\pm26\,\ms$, which leads to the total observed mass 
$m_{obs}\approx380\,\ms$. This value corresponds to about $1/3$ of the estimate given by
\citet{LK2002}. The values for FSR\,1555 are $m_{MS}=123\pm9\,\ms$, $m_{evol}=139\pm16\,\ms$,
and $m_{obs}\approx260\,\ms$. Since FSR\,1555 and LK\,1 are approximately at the same distance
from the Sun, the latter appears to be somewhat more massive than the former. 

With the more probable solution (2\,Gyr) for FSR\,1521, the total number of stars present in
$R\la5.2\arcmin$ CMD is $n_{tot}\approx161$, of which $\approx18$ are in the red clump
and $\approx111$ in the MSTO. Thus, for a MSTO mass of $\approx1.63\,\ms$, the observed
mass of FSR\,1521 is $m_{obs}\approx262\,\ms$, of which $\approx81\,\ms$ correspond to the
evolved stars, and $\approx29\,\ms$ are stored in the red clump. Interestingly, while the observed 
mass of FSR\,1521 is similar to that of FSR\,1555, the evolved mass is somewhat lower. Since both
OCs are at comparable distances, this difference is consistent with the relative ages. 

The fact that LK\,1, FSR\,1521, and FSR\,155 present similar observed masses as LK\,10, suggest 
that they all might be as massive as the latter OC.

Given the accuracy of the isochrone best-fit for old OCs provided by near-infrared
decontaminated CMDs, e.g. Fig.~\ref{fig8} and related discussions, the observed masses
are expected to be representative. Of course, deeper near-infrared photometry coupled to
similar methods as the present one, and/or including proper motion filtering, would
produce more constrained results.

\section{Discussion}
\label{Discus}

With the analyses of the preceding sections we have gathered important clues to establish that the
objects dealt with in this paper are Gyr-class OCs, or older. We also derived representative,
i.e. constrained by means of the isochrone fit (Sect.~\ref{age}), fundamental 
and structural parameters, most of which have been derived for the first time. We use these 
parameters to put the clusters into perspective, by comparing some of their properties with those 
of a set of well-studied OCs.

As reference we take the nearby OCs with ages in the range $70-7\,000$\,Myr and masses within
$400-5\,300$\,\ms\ studied by \citet{DetAnalOCs}, together with the young OCs NGC\,6611 (\citealt{N6611})
and NGC\,4755 (\citealt{N4755}). The reference clusters are distinguished according to total mass (higher 
or lower than 1\,000\,\ms). \citet{DetAnalOCs} discuss parameter correlations in the reference sample.

\begin{figure}
\resizebox{\hsize}{!}{\includegraphics{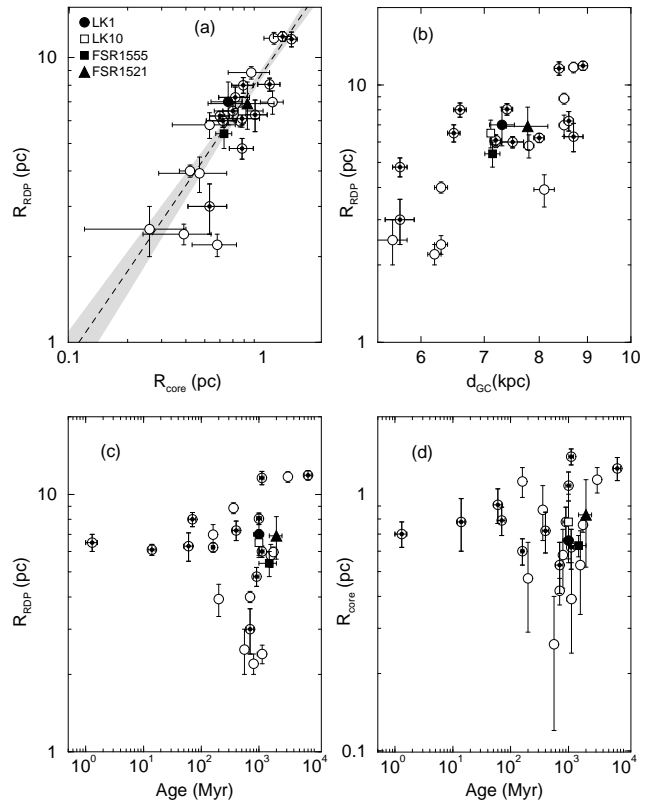}}
\caption[]{Relations involving OC structural and fundamental parameters. Circles: reference OCs. 
Dotted circles: massive ($>1\,000\,\ms$) OCs. We caution that LK\,10 may be a post-core collapse
open cluster. In this figure we adopted the core radius derived from the King-like fit. }
\label{fig10}
\end{figure}

As shown in panel (a) of Fig.~\ref{fig10}, the core and cluster radii of the reference OCs are related 
by $\rl=(8.9\pm0.3)\times R_{\rm core}^{(1.0\pm0.1)}$, which suggests a similar scaling for both kinds of 
radii, at least for the radii ranges $\rm 0.2\la\rc(pc)\la1.5$ and $\rm 2\la\rl(pc)\la15$. LK\,1, LK\,10,
FSR\,1521, and FSR\,1555 fit tightly in the relation. They also appear to follow the trend of increasing 
cluster radii with Galactocentric distance (panel b). This kind of dependence was previously suggested by, 
e.g. \citet{Lynga82}. Part of this relation may be primordial, in the sense that the higher molecular gas 
density in central Galactic regions may have produced clusters with small radii, as suggested by 
\citet{vdB91} to explain the increase of globular cluster radii with Galactocentric distance. After 
formation, mass loss associated with stellar and dynamical evolution (such as mass segregation and 
evaporation), together with tidal interactions with the Galactic potential and giant molecular clouds, 
also contribute to the depletion of star clusters, especially the low-mass and centrally located ones
(Sect.~\ref{Intro}). A similar dependence on Galactocentric distance for \rc\ is implied by the
data shown in panel (a).

In panels (c) and (d) of Fig.~\ref{fig10} we compare the presently derived cluster and core radii 
with those of the reference sample in terms of age. LK\,1, LK\,10, FSR\,1521, and FSR\,1555 have 
core and cluster radii similar to those measured in the reference OCs of equivalent age. 

\begin{figure}
\resizebox{\hsize}{!}{\includegraphics{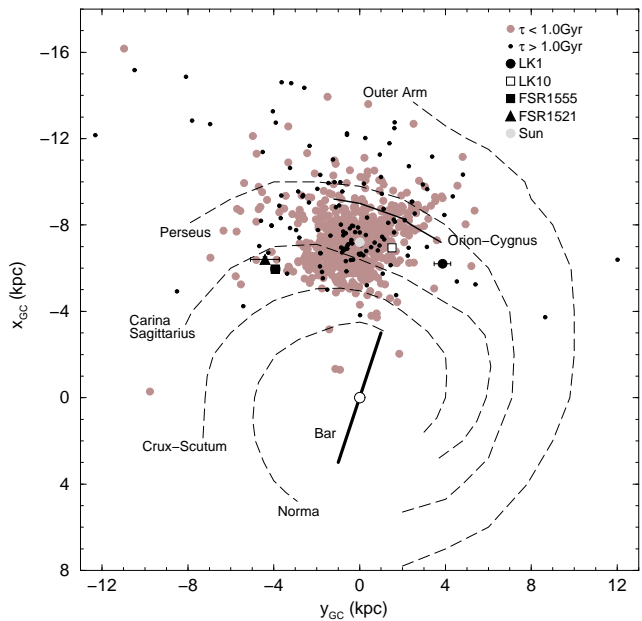}}
\caption[]{Spatial distribution of the present star clusters compared to the WEBDA OCs with 
ages younger (gray circles) and older than 1\,Gyr (black dots). Clusters are overplotted on 
a schematic projection of the Galaxy, as seen from the North pole, with 7.2\,kpc as the Sun's 
distance to the Galactic centre. Main structures are identified.}
\label{fig11}
\end{figure}

We show in Fig.~\ref{fig11} the spatial distribution of LK\,1, LK\,10, FSR\,1521, and FSR\,1555, 
as they lay in the Galactic plane. The spiral arm structure of the Milky Way is based on 
\citet{GalStr} and \citet{DrimSper01}, as derived from HII regions, and molecular clouds (e.g. 
\citealt{Russeil03}). The Galactic bar is shown with an orientation of 14\degr\ and 6\,kpc in 
total length (\citealt{Freuden98}; \citealt{Vallee05}). The present OCs are compared to the spatial 
distribution of the OCs with known age given in the WEBDA database. For comparison purposes we consider
two age groups, clusters younger and older than 1\,Gyr (the old OCs listed in Table~\ref{tab1} 
are merged into this group). As expected, old OCs are found preferentially outside the Solar circle, 
while the inner Galaxy contains few OCs so far detected. Besides, because of the presence of bright
stars, young OCs can be detected farther than the old ones, especially towards the central region.
As discussed in \citet{DiskProp}, central
directions farther than $\approx2$\,kpc begin to be critically affected by incompleteness (due to 
crowding and high backround levels) and enhanced disruption rates, while the drop in the number of 
OCs towards the opposite direction appears to be mostly real. Also, all directions show depletion
in the number of detected OCs farther than $\approx2$\,kpc. Considering the overall distribution,
the pairs LK\,1 and LK\,10, and FSR\,1521 and FSR\,1555 probe basically unexplored regions,
in opposite directions. FSR\,1555, and especially FSR\,1521, are located close to the 
Carina-Sagittarius Arm. LK\,1 and LK\,10 are found between the Carina-Sagittarius and Orion-Cygnus 
arms. 

\section{Summary and conclusions}
\label{Conclu}

Accurate fundamental and structural parameters of old star clusters are important for 
reasons that range from the completeness of the open cluster parameter space to the determination
of the efficiency of cluster dissolution mechanisms. Taken together, the results of the present 
paper, and those of previous works by our group (Table~\ref{tab1}), add to a total of 
18 old (age $\ga0.8$\,Gyr) open clusters with reliable\footnote{In the context discussed
in Sect.~\ref{age}, with uncertainties propagated from the best-fitting 2MASS isochrones.} 
parameters derived, for the first time for most of them. New findings and the age-determination
within a reasonable confidence level are important as well to improve the statistical coverage 
of the open cluster databases, especially for the definition of the old-age tail of the open
cluster age-distribution function. 

In this paper we use colour-magnitude diagrams and radial density profiles to derive 
fundamental and structural parameters of the infrared open clusters LK\,1 and
LK\,10, as well as of the star cluster candidates FSR\,1521, and FSR\,1555. Our approach 
is essentially based on field-star decontaminated 2MASS photometry, which enhances cluster 
CMD evolutionary sequences, and produces more constrained parameters.

We present consistent evidence, in the form of CMD morphology, statistical tests, structural 
parameters, and comparison with nearby OCs, that the objects are Gyr-class open clusters. With 
absorptions in the range $3.4\le\aV\le8.9$, the objects are well suited for the 2MASS photometry. 
They are located at $\ds\approx4$\,kpc from the Sun, except for the relatively nearby LK\,10, 
which is at $\ds\approx1.4$\,kpc. 
The 4 clusters are within $\approx0.6$\,kpc from the Solar circle. The old ages of LK\,1 and LK\,10 
(and the distance of the former) preclude a common origin with the Cygnus Association. With a mass 
function slope ($\chi=2.4\pm0.4$) somewhat steeper than Salpeter's ($\chi=1.35$), LK\,10 is 
relatively massive, with a total mass within $1360\le m(\ms)\le4400$. The remaining clusters
present CMD morphology and observed mass (within $260\le m(\ms)\le380$) similar to those of 
LK\,10, which suggests that they may be old clusters as massive as LK\,10. This is consistent 
especially with the relatively well-populated red clumps. 

Structurally, LK\,1, FSR\,1521, and FSR\,1555 are characterised by core and cluster radii 
similar to those of a sample of nearby open clusters of comparable age. This is consistent 
with the fact that, because they are relatively close to the Solar circle, they suffer similar 
tidal stresses. LK\,10, on the other hand, presents evidence of post-core collapse, a structure
already observed in other old OCs, such as NGC\,3960 (\citealt{N3960}).

Finally, we note that systematic searches in catalogues of star cluster candidates, coupled
with an efficient field-star decontamination algorithm, can provide important additions
the the known population of open clusters, old ones in particular. 

\section*{Acknowledgements}
The anonymous referee is acknowledged for a thorough reading of the original
manuscript and for valuable suggestions that improved the paper.
This publication makes use of data products from the Two Micron All Sky Survey, which
is a joint project of the University of Massachusetts and the Infrared Processing and
Analysis Centre/California Institute of Technology, funded by the National Aeronautics
and Space Administration and the National Science Foundation. This research has made 
use of the WEBDA database, operated at the Institute for Astronomy of the University
of Vienna. We acknowledge support from the Brazilian Institution CNPq.

\label{lastpage}

\begin{thebibliography}{}

\bibitem[\protect\citeauthoryear{Alessi, Moitinho \& Dias}{2003}]{AMD03}
   Alessi, B.S., Moitinho, A. \& Dias, W. S. 2003, A\&A, 410, 565

\bibitem[\protect\citeauthoryear{Baumgardt \& Makino}{2003}]{BM03}
   Baumgardt, H. \& Makino, J. 2003, MNRAS, 340, 227

\bibitem[\protect\citeauthoryear{van den Bergh}{1957}]{vdB57}
   van den Bergh, S. 1957, ApJ, 125, 445
   
\bibitem[\protect\citeauthoryear{van den Bergh, Morbey \& Pazder}{1991}]{vdB91}
   van den Bergh, S., Morbey, C. \& Pazder, J. 1991, ApJ, 375, 594
      
\bibitem[\protect\citeauthoryear{Bessel \& Brett}{1988}]{BesBret88}
   Bessel, M.S. \& Brett, J.M. 1988, PASP, 100, 1134
   
\bibitem[\protect\citeauthoryear{Bica, Bonatto \& Dutra}{2003}]{BBD2003}
   Bica, E., Bonatto, C. \& Dutra, C. 2003, A\&A, 405, 991

\bibitem[\protect\citeauthoryear{Bica et al.}{2006}]{GCProp}
   Bica, E., Bonatto, C., Barbuy, B. \& Ortolani, S. 2006, A\&A, 450, 105

\bibitem[\protect\citeauthoryear{Bica, Bonatto \& Blumberg}{2006}]{FaintOCs}
   Bica, E., Bonatto, C. \& Blumberg, R. 2006, A\&A, 460, 83

\bibitem[\protect\citeauthoryear{Bica et al.}{2007}]{FSR584}
   Bica, E., Bonatto, C., Ortolani, S. \& Barbuy, B. 2007, A\&A, 472, 483

\bibitem[\protect\citeauthoryear{Bica \& Bonatto}{2008}]{F1603}
   Bica, E. \& Bonatto, C. 2008, MNRAS, 384, 1733

\bibitem[\protect\citeauthoryear{Bica, Bonatto \& Camargo}{2008}]{ProbFSR}
   Bica, E., Bonatto, C. \& Camargo, D. 2008, MNRAS, 385, 349

\bibitem[\protect\citeauthoryear{Bonatto, Bica \& Girardi}{2004}]{TheoretIsoc}
   Bonatto, C., Bica, E. \& Girardi, L. 2004, A\&A, 415, 571

\bibitem[\protect\citeauthoryear{Bonatto, Bica \& Santos Jr.}{2005}]{N188}
   Bonatto, C., Bica, E. \&  Santos Jr., J.F.C. 2005, A\&A, 433, 917.

\bibitem[\protect\citeauthoryear{Bonatto \& Bica}{2005}]{DetAnalOCs}
   Bonatto, C. \&  Bica, E. 2005, A\&A, 437, 483

\bibitem[\protect\citeauthoryear{Bonatto, Santos Jr. \& Bica}{2006}]{N6611}
   Bonatto, C., Santos Jr., J.F.C. \& Bica, E. 2006, A\&A, 445, 567

\bibitem[\protect\citeauthoryear{Bonatto et al.}{2006a}]{DiskProp}
   Bonatto, C., Kerber, L.O., Bica, E. \& Santiago, B.X. 2006a, A\&A, 446, 121

\bibitem[\protect\citeauthoryear{Bonatto et al.}{2006b}]{N4755}
   Bonatto, C., Bica, E., Ortolani, S. \& Barbuy, B. 2006b, A\&A, 453, 121
   
\bibitem[\protect\citeauthoryear{Bonatto \& Bica}{2006}]{N3960}
   Bonatto, C. \& Bica, E. 2006, A\&A, 455, 931

\bibitem[\protect\citeauthoryear{Bonatto \& Bica}{2007a}]{BB07}
   Bonatto, C. \& Bica, E. 2007a, MNRAS, 377, 1301

\bibitem[\protect\citeauthoryear{Bonatto et al.}{2007}]{FSR1767}
   Bonatto, C., Bica, E., Ortolani, S. \& Barbuy, B. 2007, MNRAS, 381, L45

\bibitem[\protect\citeauthoryear{Bonatto \& Bica}{2007b}]{OldOCs}
   Bonatto, C. \& Bica, E. 2007b, A\&A, 473, 445
   
\bibitem[\protect\citeauthoryear{Bonatto \& Bica}{2008a}]{StrucPar}
   Bonatto, C. \& Bica, E. 2008a, A\&A, 477, 829

\bibitem[\protect\citeauthoryear{Bonatto \& Bica}{2008b}]{Pap11GCs}
   Bonatto, C. \& Bica, E. 2008b, A\&A, 479, 741
   
\bibitem[\protect\citeauthoryear{Bonatto \& Bica}{2008c}]{AntiCent}
   Bonatto, C. \& Bica, E. 2008c, A\&A, 485, 81

\bibitem[\protect\citeauthoryear{Bonatto, Bica \& Santos Jr.}{2008}]{PlaNeb}
   Bonatto, C., Bica, E. \& Santos Jr., J.F.C. 2008, MNRAS, 386, 324
   
\bibitem[\protect\citeauthoryear{Bonatto \& Bica}{2008d}]{Cz23}
   Bonatto, C. \& Bica, E. 2008d, A\&A, accepted (astro-ph/0809.2492)

\bibitem[\protect\citeauthoryear{Cardelli, Clayton \& Mathis}{1989}]{Cardelli89}
   Cardelli, J.A., Clayton, G.C. \& Mathis, J.S. 1989, ApJ, 345, 245
   
\bibitem[\protect\citeauthoryear{Drimmel \& Spergel}{2001}]{DrimSper01}
   Drimmel, R., \& Spergel, D.N. 2001, ApJ, 556, 181
   
\bibitem[\protect\citeauthoryear{Le Duigou \& Kn\"odlseder}{2002}]{LK2002}
   Le Duigou, J.M.  \& Kn\"odlseder, J. 2002, A\&A, 392, 869
   
\bibitem[\protect\citeauthoryear{Dutra \& Bica}{2001}]{DB2001}
   Dutra, C.M. \& Bica, E. 2001, A\&A, 376, 434

\bibitem[\protect\citeauthoryear{Dutra, Santiago \& Bica}{2002}]{DSB2002}
   Dutra, C.M., Santiago, B.X. \& Bica, E. 2002, A\&A, 383, 219

\bibitem[\protect\citeauthoryear{Eisenhauer et al.}{2003}]{Eisen03}
   Eisenhauer, F., Sch\"odel, R, Genzel, R., Ott, T., Tecza, M., Abuter, R.,
   Eckart, A. \& Alexander, T. 2003, ApJ, 597, L121

\bibitem[\protect\citeauthoryear{Eisenhauer et al.}{2005}]{Eisen05}
   Eisenhauer, F., Genzel, R., Alexander, T. et al. 2005, ApJ, 628, 246

\bibitem[\protect\citeauthoryear{Elson, Fall \& Freeman}{1987}]{EFF87}
   Elson, R.A.W., Fall, S.M. \& Freeman, K.C. 1987, ApJ, 323, 54
   
\bibitem[\protect\citeauthoryear{Freudenreich}{1998}]{Freuden98}
   Freudenreich, H.T. 1998, ApJ, 492, 495
   
\bibitem[\protect\citeauthoryear{Friel}{1995}]{Friel95}
   Friel, E.D. 1995, ARA\&A 1995, 33, 381
   
\bibitem[\protect\citeauthoryear{Froebrich, Scholz \& Raftery}{2007}]{FSRcat}
   Froebrich, D., Scholz, A. \& Raftery, C.L. 2007, MNRAS, 374, 399
   
\bibitem[\protect\citeauthoryear{Froebrich, Meusinger \& Scholz}{2007}]{FSR1735}
   Froebrich, D., Meusinger, H. \& Scholz, A. 2007, MNRAS, 377, L54
   
\bibitem[\protect\citeauthoryear{Froebrich, Meusinger \& Davis}{2007}]{FSR190}
   Froebrich, D., Meusinger, H. \& Davis, C.J. 2008, MNRAS, 383, L45

\bibitem[\protect\citeauthoryear{Girardi et al.}{2002}]{Girardi2002}
   Girardi, L., Bertelli, G., Bressan, A., et al. 2002, A\&A, 391, 195  
  
\bibitem[\protect\citeauthoryear{Goodwin \& Bastian}{2006}]{GoBa06}
   Goodwin, S.P. \& Bastian, N. 2006, MNRAS, 373, 752
   
\bibitem[\protect\citeauthoryear{von Hoerner}{1958}]{vHoerner58}
   von Hoerner, S. 1958, Astrophys., 44, 221

\bibitem[\protect\citeauthoryear{Hurley \& Tout}{1998}]{HT98}
   Hurley, J. \& Tout, A.A. 1998, MNRAS, 300, 977
   
\bibitem[\protect\citeauthoryear{Janes \& Phelps}{1994}]{JP94}
   Janes, K.A. \& Phelps, R.L. 1994, AJ, 108, 1773

\bibitem[\protect\citeauthoryear{Kerber et al.}{2002}]{Kerber02}
   Kerber, L.O., Santiago, B.X., Castro, R. \& Valls-Gabaud, D. 2002, A\&A, 390, 121

\bibitem[\protect\citeauthoryear{Khalisi, Amaro-Seoane \& Spurzem}{2007}]{Khalisi07}
   Khalisi, E., Amaro-Seoane, P. \& Spurzem, R. 2007, MNRAS, 374, 703

\bibitem[\protect\citeauthoryear{King}{1962}]{King1962}
   King, I. 1962, AJ, 67, 471

\bibitem[\protect\citeauthoryear{King}{1966}]{King66}
   King, I. 1966, AJ, 71, 64
   
\bibitem[\protect\citeauthoryear{Kroupa}{2001}]{Kroupa2001}
   Kroupa, P. 2001, MNRAS, 322, 231
   
\bibitem[\protect\citeauthoryear{Lamers \& Gieles}{2006}]{LG06}
   Lamers, H.J.G.L.M. \& Gieles, M. 2006, A\&AL, 455, 17
   
\bibitem[\protect\citeauthoryear{Lyng\aa}{1982}]{Lynga82}
   Lyng\aa, G. 1982, A\&A, 109, 213
   
\bibitem[\protect\citeauthoryear{Marigo et al.}{2008}]{Marigo08}
   Marigo, P., Girardi, L., Bressan, A., Groenewegen, M.A.T., Silva, L.
   \& Granato, G.L. 2008, A\&A, 482, 883
   
\bibitem[\protect\citeauthoryear{Massey, Johnson \& DeGioia-Eastwood}{1995}]{Massey95}
   Massey, P., Johnson, K.E. \& DeGioia-Eastwood, K. 1995, ApJ, 454, 151
   
\bibitem[\protect\citeauthoryear{Mercer et al.}{2005}]{Mercer05}
   Mercer, E.P., Clemens, D.P., Meade, M.R., Babler, B.L., Indebetouw, R.,
   Whitney, B.A., Watson, C., Wolfire, M.G. et al. 2005, ApPJ, 6335,  560
   
\bibitem[\protect\citeauthoryear{Momany et al.}{2006}]{GalStr}
   Momany, Y., Zaggia, S., Gilmore, G., Piotto, G., Carraro, G., Bedin,
   L.R. \& de Angeli, F. 2006, A\&A, 451, 515
   
\bibitem[\protect\citeauthoryear{Momany et al.}{2008}]{FSR1415}
   Momany, Y., Bica, E., Barbuy, B., Ortolani, S. \& Bonatto, C. 
   MNRAS, 2008, submitted
   
\bibitem[\protect\citeauthoryear{Naylor \& Jeffries}{2006}]{NJ06}
   Naylor, T. \& Jeffries, R.D. 2006, MNRAS, 373, 1251

\bibitem[\protect\citeauthoryear{Nishiyama et al.}{2006}]{Nishiyama06}
   Nishiyama, S., Nagata, T., Sato, S. et al. 2006, ApJ, 647, 1093
   
\bibitem[\protect\citeauthoryear{Oort}{1958}]{Oort58}
   Oort, J.H. 1958, in {\em Ricerche Astronomiche}, 5, 415, Specola Vaticana,
   Proc. of a Conference at Vatican Observatory, Castel Gandolfo, May 20-28, 1957,
   ed. D.J.K. O'Connell

\bibitem[\protect\citeauthoryear{Ortolani et al.}{2005}]{OBB05a}
   Ortolani, S., Bica, E., Barbuy, B. \& Zocalli, M. 2005, A\&A, 439, 1135
   
\bibitem[\protect\citeauthoryear{Ortolani, Bica \& Barbuy}{2005}]{OBB05b}
   Ortolani, S., Bica, E. \& Barbuy, B. 2005, A\&A, 437, 531

\bibitem[\protect\citeauthoryear{Pavani \& Bica}{2007}]{PB07}
   Pavani, D.N. \& Bica, E. 2007, MNRAS, 468, 139
   
\bibitem[\protect\citeauthoryear{Piskunov et al.}{2007}]{Piskunov07}
   Piskunov, A.E., Kharchenko, N.V., R\"oser, S., Schilbach, E. \& Scholz, R.-D.
   2007, A\&A, 445, 545
   
\bibitem[\protect\citeauthoryear{Portegies Zwart et al.}{2002}]{Portegies02}
   Portegies Zwart, S.F., Makino, J., McMillan, S.L.W. \& Hut, P.
   2002, ApJ, 565, 265
   
\bibitem[\protect\citeauthoryear{Richer et al.}{2008}]{Richer08}
   Richer, H.B., Dotter, A., Hurley, J., Anderson, J., King, I., Davis, S., Fahlman, G.G.
   Hansen, B.M.S. et al. 2008, AJ, 135, 2141
   
\bibitem[\protect\citeauthoryear{Russeil}{2003}]{Russeil03}
   Russeil, D. 2003, A\&A, 397, 133
   
\bibitem[\protect\citeauthoryear{Salaris, Weiss \& Percival}{2004}]{Salaris04}
   Salaris, M., Weiss, A. \& Percival, S.M. 2004,A\&A, 422, 217

\bibitem[\protect\citeauthoryear{Salpeter}{1955}]{Salpeter55}
   Salpeter, E. 1955, ApJ, 121, 161

\bibitem[\protect\citeauthoryear{Skrutskie et al.}{1997}]{2mass1997}
   Skrutskie, M., Schneider, S.E., Stiening, R., et al. 1997, in {\it The Impact
   of Large Scale Near-IR Sky Surveys}, ed. Garzon et al., Kluwer (Netherlands), 210, 187
   
\bibitem[\protect\citeauthoryear{Spitzer}{1958}]{Spitzer58}
   Spitzer, L. 1958, ApJ 127, 17
   
\bibitem[\protect\citeauthoryear{Trager, King \& Djorgovski}{1995}]{TKD95}
   Trager, S.C., King, I.R. \& Djorgovski, S. 1995, AJ, 109, 218

\bibitem[\protect\citeauthoryear{Upgren, Mesrobian \& Kerridge}{1972}]{Upgren72}
   Upgren, A.R., Mesrobian, W.S. \& Kerridge, S.J. 1972, AJ, 77, 74
   
\bibitem[\protect\citeauthoryear{Vall\'ee}{2005}]{Vallee05}
   Vall\'ee, J.P. 2005, AJ, 130, 569

\bibitem[\protect\citeauthoryear{Wilson}{1975}]{Wilson75}
   Wilson, C.P. 1975, AJ, 80, 175

\end{thebibliography}
\end{document}